\documentclass{emulateapj}

\def\myputfigure#1#2#3#4#5%
{\vskip#5pt\makebox[0pt]{\hskip#2in
\includegraphics[width=#3\textwidth]{#1}}\vskip#4pt\hfill}

\newcommand\lsim{\mathrel{\rlap{\lower4pt\hbox{\hskip1pt$\sim$}}
        \raise1pt\hbox{$<$}}}
\newcommand\gsim{\mathrel{\rlap{\lower4pt\hbox{\hskip1pt$\sim$}}
        \raise1pt\hbox{$>$}}}

\newcommand{\avenf}{\bar{x}_{\rm HI}}

\newcommand{\deltakvec}{\delta({\bf k})}
\newcommand{\deltaxvec}{\delta({\bf x})}
\newcommand{\sigsq}{\sigma^2(M)}
\newcommand{\sig}{\sigma(M)}
\newcommand{\delc}{\delta_c(M,z)}
\newcommand{\Msun}{M_\odot}
\newcommand{\Tvir}{T_{\rm vir}}
\newcommand{\fcoll}{f_{\rm coll}({\bf x_1}, M, z)}
\newcommand{\deltanlxvec}{\delta_{\rm nl}({\bf x_1}, z)}
\newcommand{\Tcmb}{T_\gamma}
\newcommand{\delT}{\delta T_b}

\newcommand{\delNL}{\delta_{\rm nl}}

\newenvironment{packed_enum}{
\begin{enumerate}
  \setlength{\itemsep}{1pt}
  \setlength{\parskip}{0pt}
  \setlength{\parsep}{0pt}
}{\end{enumerate}}

\begin{document}

\submitted{Submitted to the ApJ}

\title{Efficient Simulations of Early Structure Formation and Reionization}

\author{Andrei Mesinger \& Steven Furlanetto}
\affil{
Yale Center for Astronomy and Astrophysics, Yale University, New Haven, CT 06520
}
\vspace{+0.4cm}

\begin{abstract}
Detailed theoretical studies of the high-redshift universe, and especially reionization, are generally forced to rely on time-consuming N-body codes and/or approximate radiative transfer algorithms.  We present a method to construct semi-numerical ``simulations'', which can efficiently generate realizations of halo distributions and ionization maps at high redshifts.  Our procedure combines an excursion-set approach with first-order Lagrangian perturbation theory and operates directly on the linear density and velocity fields.  As such, the achievable dynamic range with our algorithm surpasses the current practical limit of N-body codes by orders of magnitude.  This is particularly significant in studies of reionization, where the dynamic range is the principal limiting factor because ionized regions reach scales of tens of comoving Mpc.  
We test our halo-finding and ionization-mapping algorithms {\it separately} against N-body simulations with radiative transfer and obtain excellent agreement.  We compute the size distributions of ionized and neutral regions in our maps.  We find even larger ionized bubbles than do purely analytic models at the same volume-weighted mean hydrogen neutral fraction, $\avenf$, especially early in reionization.  We also generate maps and power spectra of 21-cm brightness temperature fluctuations, which for the first time include corrections due to gas bulk velocities.   We find that velocities widen the tails of the temperature distributions and increase small-scale power, though these effects quickly diminish as reionization progresses.   We also include some preliminary results from a simulation run with the largest dynamic range to date: a 250 Mpc box that resolves halos with masses $M \geq 2.2\times10^8 \Msun$.  We show that accurately modeling the late stages of reionization, $\avenf \lsim 0.5$, requires such large scales. The speed and dynamic range provided by our semi-numerical approach will be extremely useful in the modeling of early structure formation and reionization.
\end{abstract}
\vspace{+0.5cm}

\section{Introduction}
\label{sec:intro}

Accurately modeling the formation of bound structures is invaluable for understanding any process in the early universe.  Reionization, the epoch when radiation from early generations of astrophysical objects managed to ionize the intergalactic medium (IGM), is particularly sensitive to the distribution of collapsed structure.  Current observations 
paint a complex picture
of the reionization epoch \citep{MH04, WL04_nf, Fan06, MH06, MR04, FZH06, MR06, Page06, Kashikawa06, Totani06}.  The next generation of instruments ({\it James Webb Space Telescope}; 21-cm instruments such as the Low Frequency Array and the Mileura Widefield Array Low-Frequency Demonstrator; CMB polarization measurements with {\it Planck}, etc.),  could potentially shed light on this poorly understood milestone.  Unfortunately, we still do not have accurate models of reionization with which to interpret these upcoming ({\it and} current) observations.

The main difficulty lies in the enormous dynamic range required. Ionized regions are expected to reach characteristic sizes of tens of comoving Mpc \citep{FZH04, FO05}, which is over seven orders of magnitude in mass larger than the pertinent cooling mass, corresponding to gas with a temperature of $T\sim10^4$ K (e.g. \citealt{Efstathiou92, TW96, Gnedin00b, SGB94}).  The required dynamic range is even larger if smaller ``minihalos" below this cooling threshold are important during reionization.  Because of the steep mass dependence of halo abundances, halos with masses close to the cooling mass could dominate the photon budget.  Hence modeling reionization requires simulation box sizes of hundreds of megaparsecs on a side, with extremely high resolution.  Attempts to overcome these obstacles have generally followed the same fundamental and well-trod path (e.g. \citealt{Gnedin00a, Razoumov02, CSW03, Sokasian03, Iliev06_sim, Zahn07, TC06}):
(1) N-body codes are run to generate halo distributions;
(2) a simple prescription is used to relate the halo mass to an ionizing efficiency;
(3) approximate methods (generally so-called ray-tracing algorithms) are used to model radiative transfer (RT) on large scales.

Even with modest halo resolution \citep{SH03} of tens of dark matter particles per halo, such schemes are computationally limited to box sizes of tens of megaparcecs, if they wish to resolve the likely cooling mass. \citet{McQuinn06} extended the mass resolution of their simulations by using a merger tree scheme to populate sub-grid scales with unresolved halos in a stochastic manner.  Such hybrid schemes are useful for extending the dynamic range, but merger trees require a number of corrections to achieve consistent mass functions (see, e.g., \citealt{sp97, bkh05}, and Fig. 1 in \citealt{McQuinn06}) and to track individual halos with redshift.  Moreover, although they are perfectly adequate for many purposes (including studying the large-scale features of reionization), they prevent one from taking full advantage of the simulation.

Aside from dynamic range, the other main limiting factor in all of the above numerical approaches is speed.  Even if the relevant scales can be resolved with N-body codes, such as may be the case in the early phase of reionization or with hybrid stochastic schemes.  The codes themselves generally take days to run on large super-computing clusters, with the approximate RT algorithms consuming a few additional days.  The computational cost of each simulation makes it difficult to explore the full range of parameter space for reionization, which is particularly large because we know so little about high-redshift galaxies.

The computational cost becomes truly prohibitive if hydrodynamics is included:  the largest such simulation of reionization performed to date spanned only $10 h^{-1}$ Mpc \citep{Sokasian03}.  Including self-consistent descriptions of galaxy formation -- even at the approximate level currently implemented in lower-redshift cosmological simulations (e.g., \citealt{SH03}) -- requires hydrodynamics, so N-body simulations of reionization are limited to semi-analytic prescriptions for star formation, feedback, etc.  It is therefore worthwhile to explore even simpler schemes.

{\it The purpose of this paper is to introduce approximate but efficient methods for generating halo distributions at high redshifts as well as for generating the associated ionization maps.}  We apply an excursion-set approach (e.g. \citealt{Bond91, LC93}) to the filtering of a realization of the linear density field and then adjust halo locations with first-order perturbation theory.
We can thus generate halo distributions at any given redshift, without explicitly including information from any higher redshifts.  This scheme is an updated form of the ``peak-patch" formalism developed and validated by \citet{BM96_algo, BM96_vali}, although it was conceived and implemented completely independently.  We then apply a similar technique to obtain the ionization field from the halo field.  
This part is similar to the schemes described in \citet{Zahn05, Zahn07}, except applied to our efficiently built halo distributions.  As such, our methods allow us to make general predictions about non-linear processes, such as structure formation and reionization, without making use of time-guzzling cosmological simulations.  The speed of our approach also allows us to explore a larger dynamic range than is possible with current cosmological simulations while preserving detailed spatial information (at least in a statistical sense), unlike purely analytic models. 

This paper is organized as follows.  In \S~\ref{sec:sim}, we introduce and test the components of our halo finding algorithm.  In \S~\ref{sec:bubble_filtering}, we introduce and test our HII bubble finding algorithm.  In \S~\ref{sec:21cm}, we use our semi-numerical scheme to generate maps and power spectra of expected 21-cm brightness temperature fluctuations throughout reionization.
In \S~\ref{sec:conc}, we summarize our key findings and present our conclusions.

Unless stated otherwise, we quote all quantities in comoving units.
We adopt the background cosmological parameters
($\Omega_\Lambda$, $\Omega_{\rm M}$, $\Omega_b$, $n$, $\sigma_8$, $H_0$)
= (0.76, 0.24, 0.0407, 1, 0.76, 72 km s$^{-1}$ Mpc$^{-1}$), consistent
with the three--year results of the {\it WMAP} satellite
\citep{Spergel06}.

\section{Semi-Numerical Simulations of Halo Properties}
\label{sec:sim}

In brief, our algorithm generates a linear density field and identifies halos within it.  Because only linear evolution is required, the algorithm is fast and flexible.  We generate 3D Monte-Carlo realizations of the linear density field on a box with sides of length $L=100$ Mpc and $N=1200^3$ grid cells.  As such, we are able to take advantage of many pre-existing tools operating on the linear density field alone.  Our method consists of the following principal steps:
\begin{packed_enum}
\item creating the linear density and velocity fields
\item filtering halos from the linear density field using the excursion-set formalism
\item adjusting halo locations using their linear-order displacements
\end{packed_enum}
Step (1) only needs to be done once for each realization, since it is independent of redshift. As mentioned above, steps (2) and (3) need only be performed on redshifts of interest, i.e. since our output at redshift $z$ is independent of any outputs at higher redshifts, there is no need for our code to ``run down'' to $z$, as is the case for N-body codes.

Our algorithm is an updated and simplified version of the ``peak-patch" algorithm of \citet{BM96_algo}; we refer the interested reader there for more detailed explanations of some steps.  A simpler version has also been used by \citet{SFM02} to study metal enrichment at high redshifts.   

We perform our semi-numerical simulations on a single desktop Mac Pro with 
two dual-core 3.00 GHz Quad Xeon processors and 16 GB of RAM.  For $N=1200^3$, step (1) takes $\sim$ 1 hour.  For a given redshift in our range of interest, specifically for $z=8.75$, steps (2) and (3) take $\sim$ 2.5 hours.  
To achieve comparable halo mass resolution (including halos with $M \gsim 10^7 \Msun$) with a minimum of $\sim500$ particles per halo \citep{SH03}, N-body codes would require a prohibitively large number of particles, $N\sim10^{12}$!  Below we describe in detail the components of our model.

\subsection{The Linear Density Field}
\label{sec:den}

Our linear density field is generated in much the same way as it is for N-body codes.  We briefly outline the procedure here.

The density field of the universe, $\deltaxvec$ $\equiv$ $ \rho({\bf x}) / \bar{\rho} - 1$, in the linear regime\footnote{In linear theory, density perturbations evolve in redshift as $\delta(z)$ = $\delta(0) D(z)$, where $D(z)$ is the linear growth factor normalized so that $D(0)$ = 1 (e.g., \citealt{Liddle96}).  Unless the redshift dependence is noted explicitly, from this point forward we will work with quantities linearly-extrapolated to $z=0$.}  is well-represented as a Gaussian random field, whose statistical properties are fully defined by its power spectrum, $\sigma^2({\bf k}) \equiv \langle |\deltakvec |^2 \rangle$.  Here, $\deltakvec$ is the Fourier transform of $\deltaxvec$, and the standard assumption of isotropy implies $\sigma^2({\bf k})=\sigma^2(k)$ while homogeneity implies that there are no density fluctuations with wavelengths larger than the box size $L = V^{1/3}$.  We use the following, standard (e.g. \citealt{BP97, Sirko05}) Fourier transform conventions:

\begin{equation}
\label{eq:deltak}
\deltakvec = \frac{V}{N} \sum \deltaxvec e^{-i {\bf k \cdot x}} ~ ,
\end{equation}
\noindent with the inverse transform being
\begin{equation}
\label{eq:deltax}
\deltaxvec = \frac{1}{V} \sum \deltakvec e^{i {\bf k \cdot x}} ~ .
\end{equation}

The discrete simulation box only permits a finite set of wavenumbers: ${\bf k} = \Delta k (i, j, k)$, where $\Delta k = 2\pi/L$ and $i, j, k$ are integers in the range (-$\sqrt[3]{N}$/2, $\sqrt[3]{N}$/2].  For each independent wavenumber,\footnote{Since $\deltaxvec$ is real-valued, only half of the $k$-modes defined above are independent.  The other half are determined by the usual Hermitian constraints for real-valued functions (see for example \citealt{HE88, BP97}).} we assign

\begin{equation}
\label{eq:deltak_prob}
\deltakvec = \sqrt{\frac{\sigma^2(k)}{2}} (a_{\bf k} + i b_{\bf k}) ~ ,
\end{equation}

\noindent where $a_{\bf k}$ and $b_{\bf k}$ are drawn from a zero-mean Gaussian distribution with unit variance.  We use the power spectrum from \citet{EH99}. Then the real-space density field, $\deltaxvec$, is obtained by performing an inverse Fourier transform on $\deltakvec$.

\subsection{The Linear Velocity Field}
\label{sec:velocity}

We construct a linear velocity field corresponding to our linear density field using the standard Zel'Dovich approximation (c.f. \citealt{ZelDovich70, Efstathiou85, Sirko05}):

\begin{eqnarray}
\label{eq:zeldovich}
{\bf x_1} &=& {\bf x} + {\bf \psi}({\bf x}) ~ ,\\
{\bf v} &\equiv& {\bf \dot{x}_1} = \dot{{\bf \psi}}({\bf x}) ~ ,\\
\nonumber \delta({\bf x}) &=& - {\bf \nabla \cdot} [(1+\delta({\bf x})){\bf \psi}({\bf x})] \\
&\approx& - {\bf \nabla \cdot \psi}({\bf x}) ~ ,
\end{eqnarray}

\noindent where ${\bf x}$ and ${\bf x_1}$ denote initial (Lagrangian) and updated (Eulerian) coordinates, respectively, ${\bf \psi}({\bf x})$ is the displacement vector, and the last equation follows from the continuity criterion, with the final approximation using linearity, $\delta({\bf x}) \ll 1$.  We note again that all units are comoving, unless stated otherwise.  From the above, one can relate the velocity mode in our simulation at redshift $z$ to the linear density field:

\begin{equation}
\label{eq:velocity}
{\bf v}({\bf k}, z) = \frac{i{\bf k}}{k^2} \dot{D}(z) \delta({\bf k}) ~ ,
\end{equation}

\noindent where for computational convenience differentiation is performed in $k$-space.

Another convenient property of this first-order Zel'Dovich approximation is that the velocity field can be decomposed into purely spatial, ${\bf v}_x({\bf x})$, and purely temporal, $v_{z}(z)$, components:
\begin{equation}
\label{eq:vel_comp}
{\bf v}({\bf x}, z) = v_{z}(z) {\bf v}_x({\bf x}) ~ , 
\end{equation}

\noindent where $v_{z}(z) = \dot{D}(z)$ and ${\bf v}_x({\bf x})$ is the inverse Fourier transform of $i{\bf k} \delta({\bf k})/k^2$.  This is computationally convenient, as we only need to compute the ${\bf v}_x({\bf x})$ field once in order to be able to scale it for all redshifts, and it also allows us to write a simple, exact expression for the {\it integrated} linear displacement field, ${\bf \Psi}$.  When eq. (\ref{eq:vel_comp}) is integrated from some large initial $z_0$ [$D(z_0) \ll D(z)$], the total displacement is just

\begin{eqnarray}
\label{eq:displacement}
\nonumber {\bf \Psi} ({\bf x}, z) &=&  [D(z)-D(z_0)] {\bf v}_x({\bf x}) \\
&\approx& D(z) {\bf v}_x({\bf x}) ~
\end{eqnarray}

\noindent  We make use of this displacement field to adjust the halo locations obtained by our filtering procedure (see \S~\ref{sec:displacement}), as well as to adjust the linear density field for our 21-cm temperature maps (see \S~\ref{sec:21cm}).

In principle, one could obtain non-linear velocities by mapping the linear overdensity to a corresponding  non-linear overdensity obtained from a spherical collapse model \citep{MW96}, and then taking the time derivative of the non-linear overdensity.  However, due to the large spread in the dynamical times of the non-linear density field, accurately capturing the time evolution is non-trivial.  Furthermore, although the non-linear density field implicitly captures the velocities of collapsing gas, mapping each pixel's linear density to its non-linear counterpart independently of other nearby pixels does not properly preserve correlations on larger scales.  Hence, we choose to use the linear density field directly in estimating velocities.  For the purposes of studying the ionization field, we are further justified in this procedure because our final ionization maps are smoothed on large scales, on which most pixels are still in the linear regime at the high redshifts of interests.
It is possible to include higher-order contributions to the Zel'Dovich approximation where necessary (e.g., \citealt{SS02}).

\subsection{Halo Filtering}
\label{sec:halo_filtering}

In standard Press-Schechter theory (PS; see e.g., \citealt{PS74, Bond91, LC93}), the halo mass function can be written as

\begin{equation}
\label{eq:dndM}
\frac{\partial n(>M, z)}{\partial M} = - \sqrt{\frac{2}{\pi}} \frac{\bar{\rho}}{M} \frac{\delta_c(z)}{\sigsq} \frac{\partial \sig}{\partial M} \exp \left[ - \frac{\delta_c^2(z)}{2 \sigsq} \right],
\end{equation}

\noindent where $n(>M, z)$ is the mean number density of halos with total mass greater than $M$, $\bar{\rho} = \Omega_{\rm M} \rho_{\rm crit}$ is the mean background matter density, $\delta_c(z) \sim 1.68/D(z)$ is the scale-free critical over--density evaluated in the case of spherically symmetric collapse \citep{Peebles80}, and

\begin{equation}
\label{eq:ps_sig}
\sigsq = \frac{1}{V} \int_{0}^{\infty} \frac{k^2 dk}{2 \pi^2} \sigma^2(k) W^2(k, M) ~ ,
\end{equation}

\noindent is the squared {\it r.m.s.} fluctuation in the mass enclosed within a region
described by the filter function, $W(k,M)$, normalized to integrate to unity.



Although the PS mass function in eq. (\ref{eq:dndM}) is in fair
agreement with simulations, especially for halos near the characteristic mass, at low redshifts it
underestimates the number of high--mass halos and overestimates the number of
low--mass halos when compared with large numerical simulations
(e.g. \citealt{Jenkins01}).  A modified expression shown to fit low-redshift simulation results more accurately (to within $\sim 10\%$) was obtained by \citet{ST99}:

\begin{equation}
\label{eq:st_dndM}
\frac{\partial n(>M,z)}{\partial M} = - \frac{\bar{\rho}}{M} \frac{\partial [\ln ~ \sig]}{\partial M} \sqrt{\frac{2}{\pi}}
A \left(1 + \frac{1}{\hat{\nu}^{2p}} \right) \hat{\nu} e^{ - \hat{\nu}^2/2},
\end{equation}

\noindent where $\hat{\nu} \equiv \sqrt{a} \delta_c(z) / \sig$,
and $a$, $p$, and $A$ are fitting parameters.  \citet{SMT01} derive this form of the mass function by including shear and
ellipticity in modeling non--linear collapse, effectively changing the
scale-free critical over--density $\delta_c(z)$, into a
function of filter scale,

\begin{equation}
\label{eq:st_delc}
\delc = \sqrt{a} \delta_c(z) \left[ ~ 1 + b \left(\frac{\sigsq}{a \delta_c^2(z)}\right)^c ~ \right].
\end{equation}

\noindent Here $b$ and $c$ are additional fitting parameters ($a$ is the same as
in eq.~\ref{eq:st_dndM}).  For the constants above, we adopt the recent values obtained by \citet{Jenkins01}, who studied a large range in redshift and mass: $a$ =
0.73, $A$ = 0.353, $p$ = 0.175, $b$ = 0.34, $c$ = 0.81.
We note, however, that the situation at high redshifts is less clear:  studies disagree on the relative accuracy of the \citet{PS74} and \citet{Jenkins01} forms \citep{Reed03,Iliev06_sim, Zahn07}.  Our algorithm can be trivially modified to accommodate other choices for the mass function; fortunately, for the purposes of the ionization maps (see \S \ref{sec:bubble_filtering}), the choice of mass function makes very little difference because all have a similar dependence on the local density \citep{FMH05}.  

The mass functions in equations (\ref{eq:dndM}) and (\ref{eq:st_dndM}) can be obtained by the standard excursion set random walk procedure.  The approach is to smooth the density field around a point, {\bf x}, on successively smaller scales starting with $M \rightarrow \infty$ [where $\sigsq \rightarrow 0$] and to identify the point as belonging to the halo with the largest $M$ such that $\delta({\bf x}, M) > \delta_c(M, z)$. If $W^2(k, M)$ is chosen to have a sharp cut-off, this procedure amounts to a random walk of $\delta({\bf x}, M)$ along the mass axis, since the {\it change} in $\delta({\bf x}, M)$ as the scale is shrunk is independent of $\delta({\bf x}, M)$ for a top-hat filter in $k$-space (see eq. \ref{eq:ps_sig}).

We perform this procedure on our realization of the linear density field by filtering the field using a real-space top-hat filter\footnote{There is a slight swindle in the current application of this formalism. The filter function is assumed to be a top-hat in {\it $k$-space} in order to facilitate the analytic random walk approach described above.  However, when the power spectrum is normalized to observations [i.e. $\sigma(R=8 h^{-1} {\rm Mpc}]$ = $\sigma_8$), the filter that is used to define the mass $M$ corresponding to $R$ is a top-hat filter in {\it real space}.  Nevertheless, it has been shown that the mass function is not very sensitive to this filter choice (Bond et al. 1991).}, starting on scales comparable to the box size and going down to grid cell scales, in logarithmic steps of width $\Delta M / M = 1.2$.\footnote{We note that \citet{MPH05} required a much smaller step size at these redshifts, $\Delta M / M \sim 0.1$, in order to produce accurate mass functions using 1D Monte-Carlo random walks.  However, here we find that we can reproduce accurate mass functions with a larger step size, since in our 3D realization of the density field, ``overstepping'' $\delta_c(M, z)$ due to a large filter step size can be compensated with a small offset in the filter center, i.e. by centering the filter in a neighboring cell.  This is the case since overstepping $\delta_c(M, z)$ means that some dense matter between the two filter scales was ``missed''.  In a 1D Monte-Carlo random walk this matter is unrecoverable; however, in a 3D realization of the density field, the missed matter will be picked up by a filter centered on a neighboring cell.}  At each filter scale, we use the scale-dependent barrier in eq. (\ref{eq:st_delc}) to mark a collapsed halo if $\delta({\bf x}, M) > \delta_c(M, z)$.  Filter scales large enough that collapsed structure is extremely unlikely, $\delta_c(M,z) > 7\sigma(M)$, are skipped \citep{MPH05}.  Since this procedure treats each cell as the center of a spherical filter, neighboring pixels are not properly placed in the same halo.  Because of this, we discount halos which overlap with previously marked halos.

As mentioned above, this algorithm is similar to the ``peak-patch" approach first introduced by \citet{BM96_algo}.  The primary differences are:  (1) we use the \citet{Jenkins01} barrier to identify halos (rather than calculating the strain tensor to account for ellipsoidal collapse), (2) we do not separately identify peaks in the density field (this step is not required given modern computing power), and (3) we use the ``full exclusion" criterion for preventing halo overlap.  \citet{BM96_vali} found that a ``binary exclusion" method in which pairs of overlapping halos are compared and eliminated was somewhat more accurate.  However, at the high redshifts of interest to us, halo overlap is rare, and we are primarily interested in the large-scale properties of the halo field, which are relatively insensitive to the details of the overlap criterion.

We also note that our halo finder is similar in spirit to the  PTHalos algorithm introduced by \citet{SS02} to generate mock galaxy surveys at low redshifts. There are two key differences.  First, at present we use 
only first-order perturbation theory to displace the particles.\footnote{We note that a similar scheme to ours has been independently created by O. Zahn (private communication).  This scheme uses a simple Press-Schechter barrier but adjusts halo locations following second-order Lagrangian perturbation theory.  However, he has found that the second-order corrections make very little difference to the map.} This limits us to higher redshifts, where velocities are smaller.  However, our algorithm does not require particles in order to resolve halos and hence can accommodate a considerably larger dynamic range than PTHalos.

\begin{figure}
\vspace{+0\baselineskip}
\myputfigure{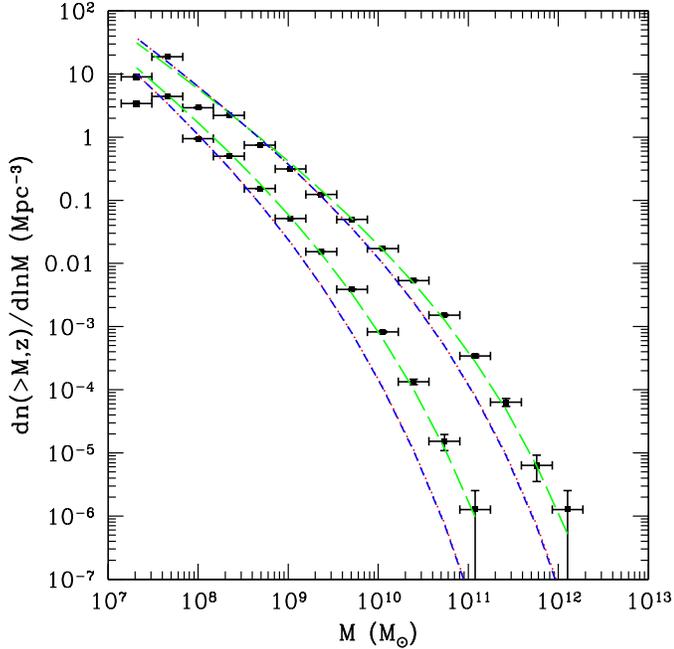}{3.3}{0.5}{.}{0.}  \figcaption{ 
Mass functions generated from our halo filtering procedure discussed in \S \ref{sec:halo_filtering} are shown as points. Dotted red curves denote PS mass functions generated by eq. (\ref{eq:dndM}); short--dashed blue curves denote extended PS conditional mass functions generated by eq. (\ref{eq:dndM}) but also taking into account the absence of density modes longer than the box size; long--dashed green curves denote mass functions generated using the Sheth-Tormen correction in eq. (\ref{eq:st_dndM}).  The upper (lower) set of curves and points correspond to redshifts of $z$=6.5 ($z$=10).  
\label{fig:dndlnm}}
\vspace{-1\baselineskip}
\end{figure}

Mass functions resulting from this procedure are shown as points in Figure \ref{fig:dndlnm}, with error bars indicating 1-$\sigma$ Poisson uncertainties and bin widths spanning our mass filter steps.
Dotted red curves denote PS mass functions generated by eq. (\ref{eq:dndM}); short--dashed blue curves denote extended PS conditional mass functions generated by eq. (\ref{eq:dndM}) but also taking into account the absence of density modes longer than the box size;
 long--dashed green curves denote mass functions generated using the Sheth-Tormen correction in eq. (\ref{eq:st_dndM}).  The upper (lower) set of curves and points correspond to redshifts of $z$=6.5 ($z$=10).  The dotted and short--dashed curves overlap at these redshifts due to our large box size ($L=100$ Mpc), so we are immune to the finite box effects pointed out by \citet{BL04}.

Fig. \ref{fig:dndlnm} shows that we obtain accurate mass functions for $M \gsim 10^8 \Msun$.
Our procedure seems to underpredict the abundance of halos with masses approaching the cell size, $M_{\rm cell} \sim 10^7 \Msun$.  However, as the Jeans mass corresponding to a gas temperature of $\sim 10^4$ K is $M_J(z\sim8)\sim 10^8 \Msun$, in subsequent calculations, we only use halos with masses greater than $M_{\rm min} = 10^8 \Msun$.  Using this $M_{\rm min}$, we match the collapse fraction obtained by integrating eq. (\ref{eq:st_dndM}) to better than $\sim10\%$.

This mass cutoff corresponds to the minimum temperature required for efficient atomic hydrogen cooling and would be the pertinent mass scale if: (1) the H$_2$ cooling channel is suppressed, e.g. due to a pervasive Lyman-Werner (LW) background, and if (2) photo-ionization feedback is ineffective at suppressing gas cooling and collapse onto higher mass halos.  While feedback at high redshifts remains poorly-constrained, both of these assumptions seem reasonable during the middle stages of reionization on which we focus.  A dissociating  LW background is likely to have established itself well before the universe is significantly ionized \citep{HRL97}.  Model-dependent empirical evidence supporting the suppression of star formation in smaller mass halos, $M \lsim M_{\rm min}$, can also be gleaned from {\it WMAP} data \citep{HB06}.  Furthermore, although early work suggested that an ionizing background could partially suppress star formation in halos with virial temperatures of $\Tvir\lsim3.6\times10^5$ K ($M \lsim 2 \times 10^9 \Msun$) \citep{TW96}, more recent studies \citep{KI00, Dijkstra04} find that at high redshifts ($z \gsim 3$), self-shielding and the increased cooling efficiency could be strong countering effects for halos with virial temperatures $\Tvir>10^4$ K.  We postpone a more detailed analysis of the reionization footprint left by photo-ionization feedback to a future work.

\subsection{Adjusting Halo Locations}
\label{sec:displacement}

Once the halo field is obtained, we use the displacement field obtained through eq. (\ref{eq:displacement}) to adjust the halo locations at each redshift.  This corrects for the enhanced halo bias in Eulerian space with respect to our filtering, which is done in Lagrangian space (i.e. using the initial locations at large $z$).   For computational convenience, we smooth the 1200$^3$ velocity field onto a coarser-grained 200$^3$ grid before adjusting halo locations.  The choice of resolution, where each cell is (100 Mpc)/200 = 0.5 Mpc on a side, is somewhat arbitrary here, and we have verified that our halo and 21-cm power spectra are unaffected by this choice. We also note that in linear theory, the mean velocity dispersion inside a (0.5 Mpc)$^3$ sphere with mean density at $z=10$ is a factor of $\sim$10 lower than the {\it r.m.s.} bulk velocity of such regions, so smoothing over smaller scale velocities appears reasonable. Furthermore, we keep in mind that our ``endproducts'' in this work are ionization and 21-cm temperature fluctuation maps, for which such ``low-resolution'' is more than adequate (compare, e.g., to N-body simulations of reionization, which typically have similar cell sizes for the radiative transfer component).

\begin{figure}
\vspace{+0\baselineskip}
\myputfigure{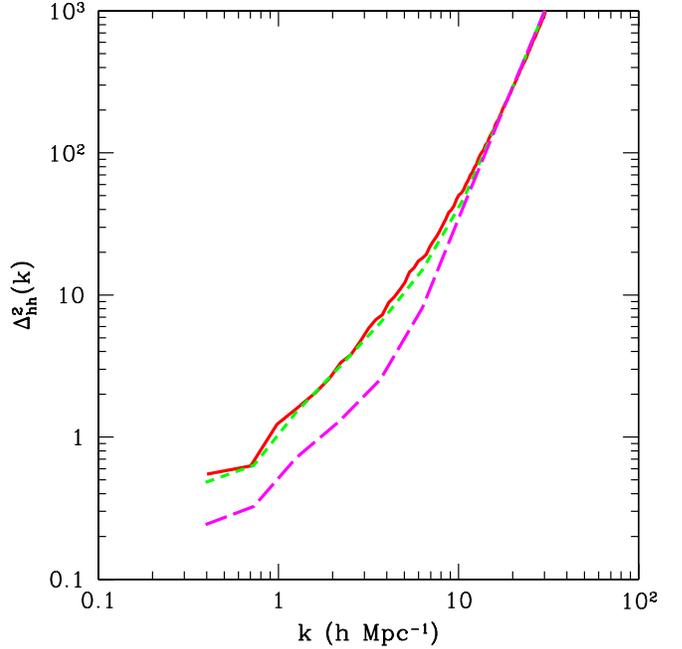}{3.3}{0.5}{.}{0.}  \figcaption{ 
Halo power spectra at $z=8.7$, with $L=20$ $h^{-1}$Mpc and cosmological parameters taken from \citet{McQuinn06}.  The solid red curve is the halo power spectrum from an N-body simulation obtained from \citet{McQuinn06} (c.f. the bottom panel of their Fig. 2). The short-dashed green and the long-dashed violet curves are obtained from our filtering procedure  with and without the halo location adjustments, respectively.
\label{fig:halo_ps}}
\vspace{-1\baselineskip}
\end{figure}

In Figure \ref{fig:halo_ps}, we plot the halo power spectrum, defined as $\Delta_{\rm hh}(k, z) = k^3/(2\pi^2 V) ~ \langle|\delta_{\rm hh}({\bf k}, z)|^2\rangle_k$, where $\delta_{\rm hh}({\bf x}, z) \equiv M_{\rm coll}({\bf x}, z)/ \langle M_{\rm coll}(z)\rangle - 1$ is the collapsed mass 
field.\footnote{We use the collapsed mass field, rather than the individual galaxies, because we calculate the power from the smoothed cells.}  The solid red curve is the halo power spectrum from a 20 $h^{-1}$ Mpc N-body simulation at $z=8.7$ obtained from \citet{McQuinn06} (c.f. the bottom panel of their Figure 2). The short-dashed green and the long-dashed violet curves are obtained from our filtering procedure (matching the assumed cosmology) with and without the halo location adjustments, respectively.  We note that ignoring the cumulative motions of halos results in an underestimate of the power of long-wavelength modes of the halo field by a factor of $\sim2$ in this case.  
The average Eulerian bias of these halos is $\sim 2$, about half of which comes from the correction from Lagrangian to Eulerian coordinates.

After the halo locations are adjusted according to linear theory, our halo power spectrum agrees almost perfectly with the simulation.  By design our procedure includes Poisson fluctuations in the halo number counts, which dominate the power spectrum at $k\gsim5$ $h$/Mpc and are lost in purely analytic estimates \citep{McQuinn06}.  We also note that both the halo mass functions and power spectra are statistical tests and hence the agreement shown here does not imply that our halo field has a one-to-one mapping with an N-body halo field sourced by identical initial conditions.  Indeed, \citet{GB94} showed that those particles located nearest initial linear density peaks are not necessarily incorporated into massive galaxies.  The ``peak particle" algorithm is less robust than our smoothing technique, but we still do not expect to recover halo masses or locations precisely.  We plan on doing a ``one-on-one'' comparison between halo fields obtained from our halo finder to those obtained from N-body codes in a future work.  However, it is certainly encouraging that the very similar ``peak-patch'' group finding formalism of \citet{BM96_algo} did very well when compared ``one-on-one'' to N-body codes at large mass scales \citep{BM96_vali}.

\begin{figure*}
\vspace{+0\baselineskip}
\includegraphics[width=0.5\textwidth]{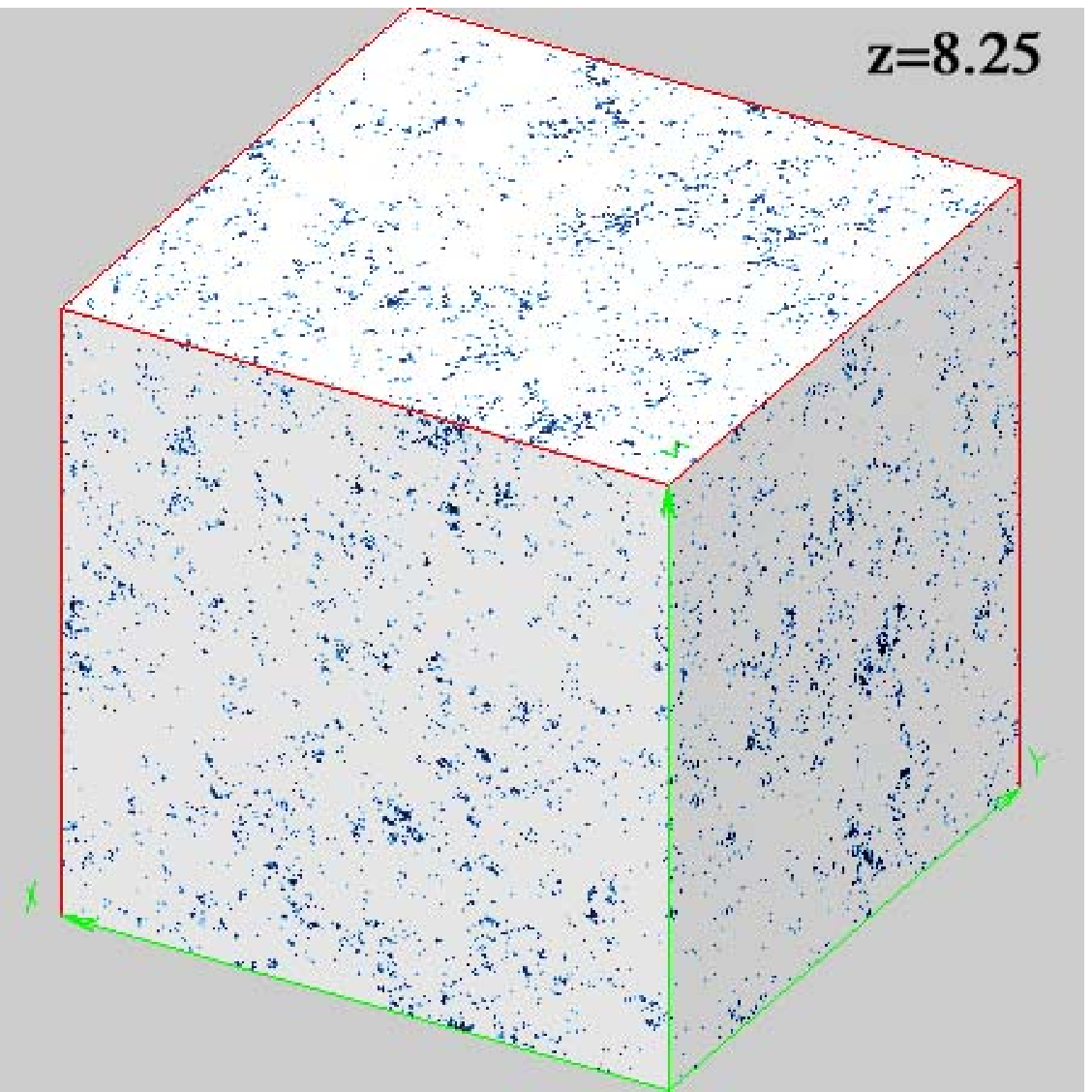}
\includegraphics[width=0.5\textwidth]{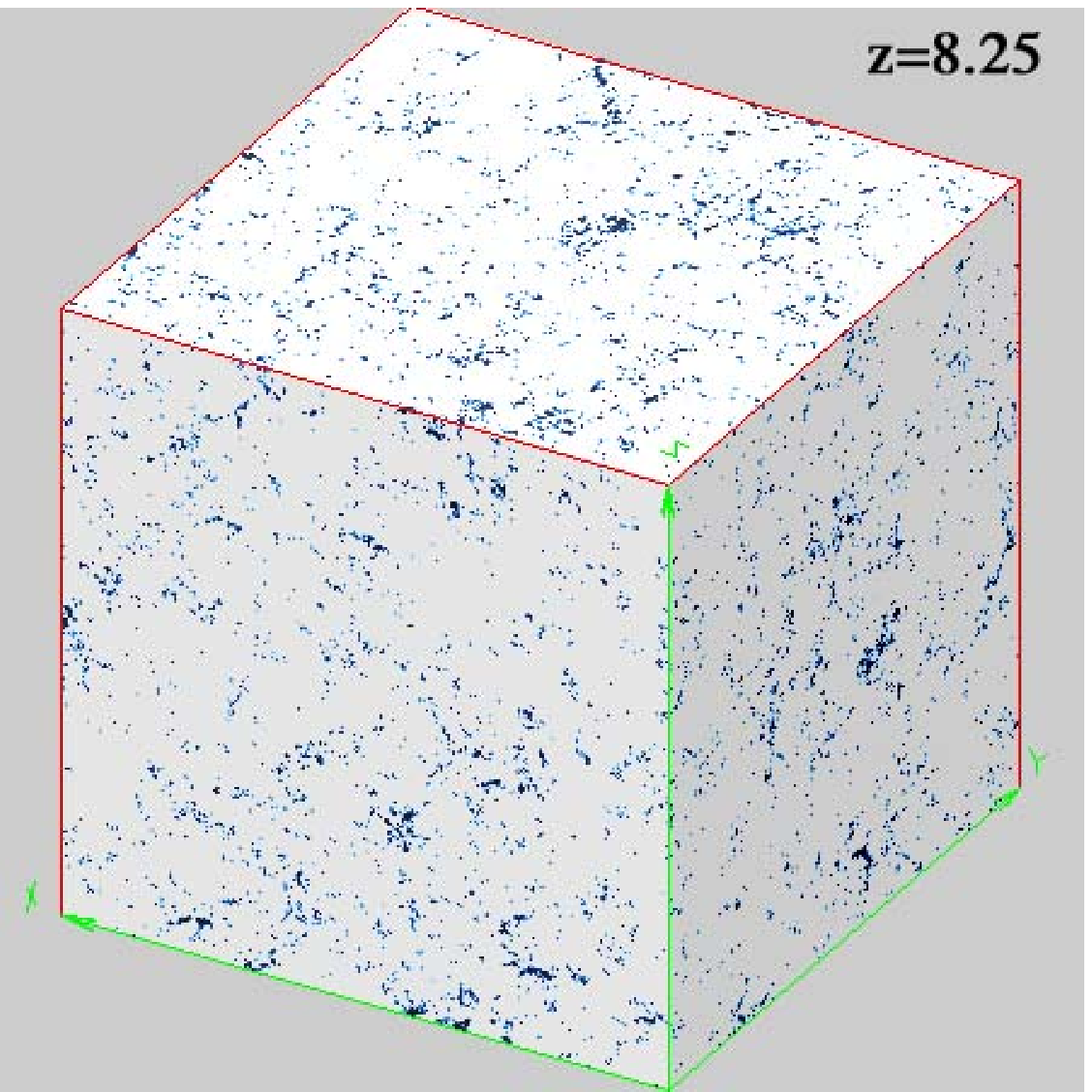}
 \figcaption{ 
Slices through the halo field from our simulation box at $z=8.25$.  The halo field is generated on a $1200^3$ grid and then mapped to a $400^3$ grid for viewing purposes.  Each slice is 100 Mpc on a side and 0.25 Mpc deep.  Collapsed halos are shown in blue.  The left panel shows the halo field directly filtered in Lagrangian space; the right panel maps the field to Eulerian space according to linear theory (see \S~\ref{sec:displacement} and eq. \ref{eq:displacement}).  The right panel corresponds to the bottom-left ($\avenf=0.53$) ionization field in Figure \ref{fig:bubble_maps}.
\label{fig:halo_slice}}
\vspace{-1\baselineskip}
\end{figure*}

In Figure \ref{fig:halo_slice} we show slices through the halo field from our simulation box at $z=8.25$, generated by the above procedure, again with ({\it right panel}) and without ({\it left panel}) the halo location adjustments.  In the figure, the halo field is mapped to a lower resolution $400^3$ grid for viewing purposes.  Each slice is 100 Mpc on a side and 0.25 Mpc deep.  Collapsed halos are shown in blue.  Visually, it is obvious that peculiar motions increase halo clustering.

\section{Generating the Ionization Field}
\label{sec:bubble_filtering}

Once the halo field is generated as described above, we can perform a similar filtering procedure (also using the excursion-set formalism) to obtain the ionization field (similar methods have been discussed by \citealt{Zahn05,Zahn07}).  The time required for this final step is a function of $\avenf$, with large $\avenf$ requiring less time than small $\avenf$.  Specifically, at $\avenf\sim0.5$ this step takes $\sim 15$ minutes to generate a 200$^3$ ionization box on our workstation.

There are two main differences between the halo filtering and the HII bubble filtering procedures: (1) HII bubbles are allowed to overlap, and (2) the excursion set barrier (the criterion for ionization) becomes, as per \citet{FHZ04}:

\begin{equation}
\label{eq:HII_barrier}
\fcoll \geq \zeta^{-1} ~ ,
\end{equation}

\noindent where $\zeta$ is some efficiency parameter and $\fcoll$ is the fraction of mass residing in collapsed halos inside a sphere of mass $M=4/3 \pi R^3 \bar{\rho} [1+\langle \deltanlxvec \rangle_R]$, with mean physical overdensity $\langle \deltanlxvec \rangle_R$, centered on Eulerian coordinate ${\bf x_1}$, at redshift $z$.

Equation~(\ref{eq:HII_barrier}) is only an approximate model and makes several simplifying assumptions about reionization.  In particular, it assumes a constant ionizing efficiency per halo and ignores spatially-dependent recombinations and radiative feedback effects.  It can easily be modified to include these effects (e.g., \citealt{FZH04_21cmtop, FMH05, FO05}), and we plan to do so in future work.  Here we present the simplest case in order to best match current RT numerical simulations.

This prescription models the ionization field as a two-phase medium, containing fully-ionized regions (which we refer to as HII bubbles) and fully-neutral regions.  This is obviously much less information than can be gleaned from a full RT simulation, which precisely tracks the ionized fraction.  However, HII bubbles are typically highly-ionized during reionization, and for many purposes (such as for 21 cm maps) this two-phase approximation is perfectly adequate.  

In order to ``find'' the HII bubbles at each redshift we smooth the halo field onto a $200^3$ grid.  Then we filter the halo field using a real-space top-hat filter, starting on scales comparable to the box size and decreasing to grid cell scales in logarithmic steps of width $\Delta M / M = 0.33$.   At each filter scale, we use the criterion in eq. (\ref{eq:HII_barrier}) to check whether the region is ionized.  If so, we flag all pixels inside that region as ionized.  We do this for all pixels and scales, regardless of whether the resulting bubble would overlap with other bubbles.  Note, therefore, that the nominal ionizing efficiency $\zeta$ that we use as an input parameter does not equal $(1-\avenf)/f_{\rm coll}$.
They typically differ by $\la 30\%$, with $\zeta f_{\rm coll} < 1-\avenf$ early in reionization and $\zeta f_{\rm coll} > 1-\avenf$ late in reionization).  Unfortunately, we thus cannot use our algorithm to self-consistently predict the time evolution of the ionized fraction (rather, that must be prescribed from some other model).  Of course, the same is true for N-body simulations, because the evolution of the ionized fraction depends on the evolving ionization efficiency of galaxies and cannot be self-consistently included in any present-day simulation.

In order to obtain the density field used in eq. (\ref{eq:HII_barrier}), $\deltanlxvec$, we use the Zel'Dovich approximation on our linear density field, $\deltaxvec$, in much the same manner as we did to adjust our halo field in \S~\ref{sec:displacement}.  Starting at some arbitrarily large initial redshift (we use $z_0=50$), we discretize our high-resolution 1200$^3$ field into ``particles'' whose mass equals that in each grid cell.  We then use the displacement field (eq. \ref{eq:displacement}) to move the particles to new locations at each redshift.  This resulting mass field is then smoothed onto our lower resolution 200$^3$ box to obtain $\deltanlxvec$.  We then recalculate the velocity field (\S~\ref{sec:velocity}) using the new densities.

\citet{Zahn07} showed that a very similar HII bubble filtering procedure performed on an N-body halo field was able to reproduce the ionization topology obtained through a ray-tracing RT algorithm fairly well.  Their algorithm differs from ours in two ways.  First, they used a slightly different barrier definition; however, this difference has only a small impact on the ionization topology.\footnote{Specifically, in order to match the physics of their simulations better, they required $\int dt \, f_{\rm coll}>\zeta^{-1}$.  However, the density modulation ends up nearly identical to our model, so the topology is almost unchanged.}  More importantly, for each filter scale at each pixel, \citet{Zahn07} flag only the center pixel as ionized if the barrier is crossed, whereas we flag the entire filtered sphere.  

\begin{figure*}
\vspace{+0\baselineskip}
{
\includegraphics[width=0.245\textwidth]{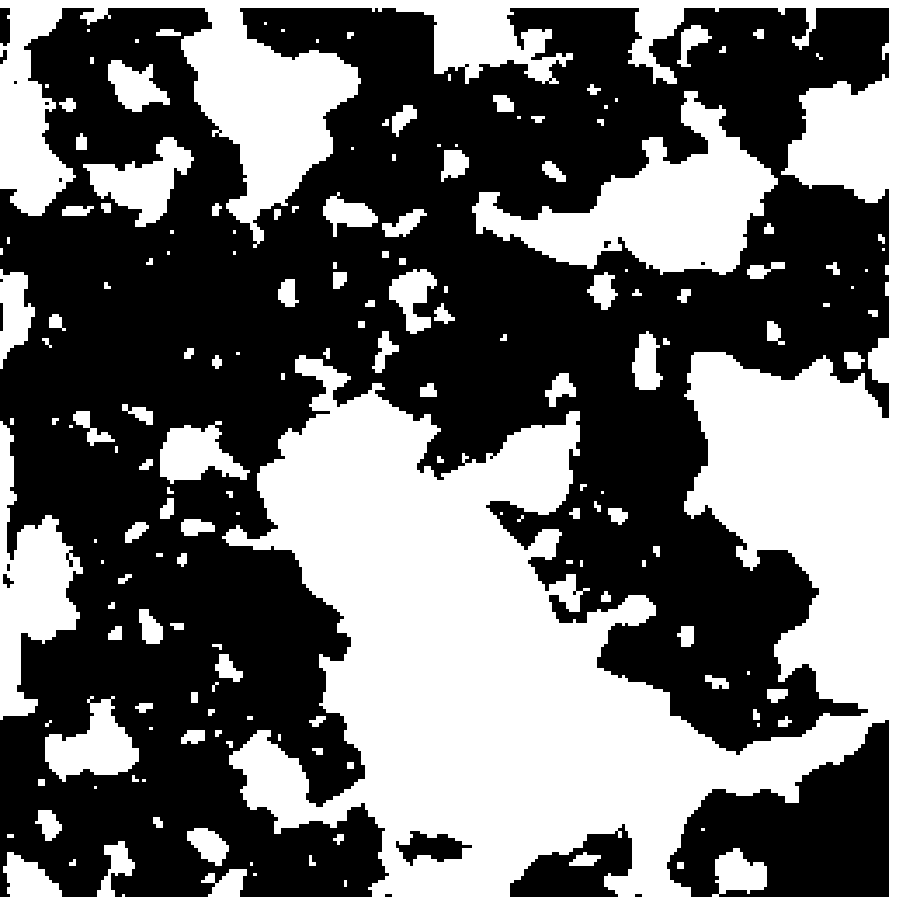}
\includegraphics[width=0.245\textwidth]{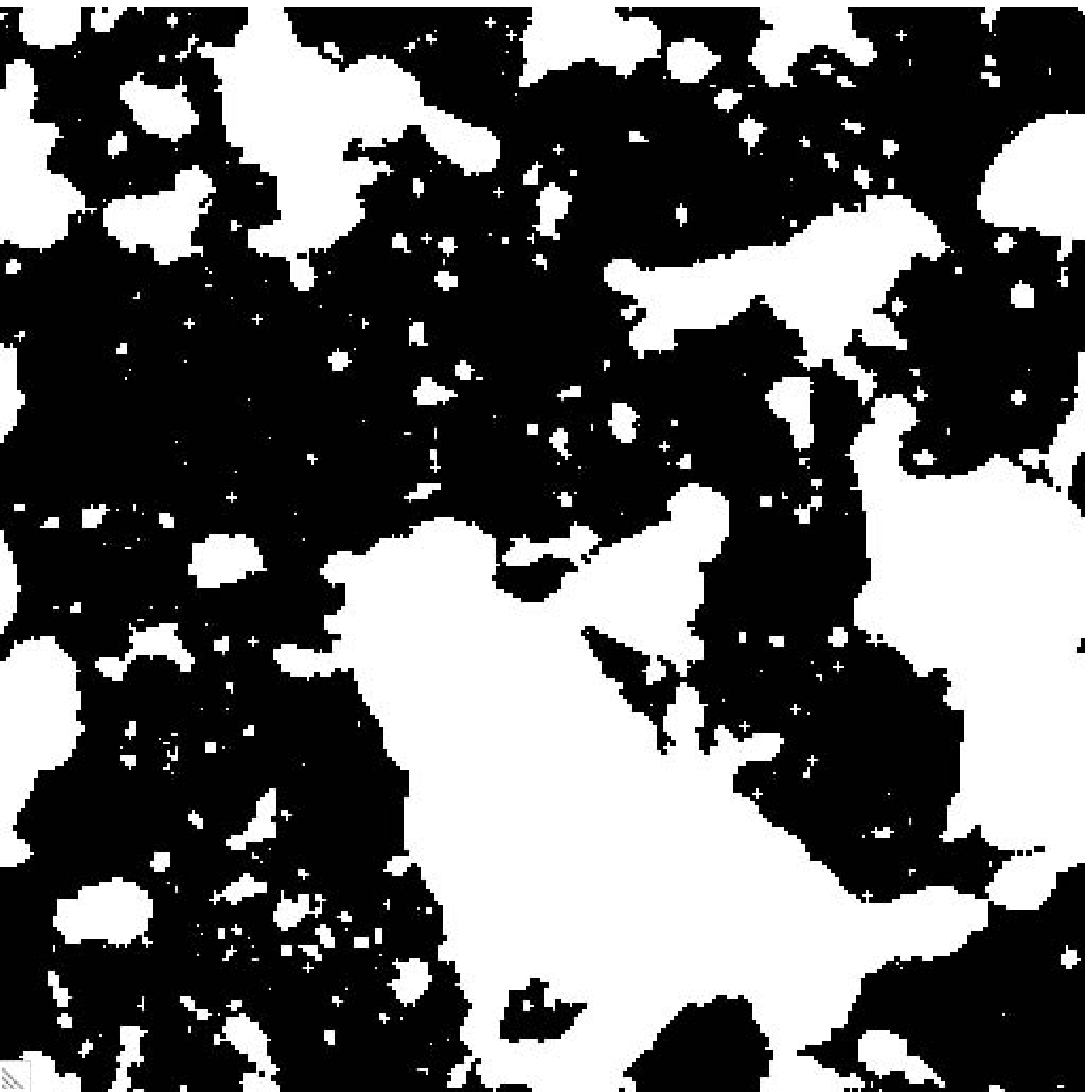}
\includegraphics[width=0.245\textwidth]{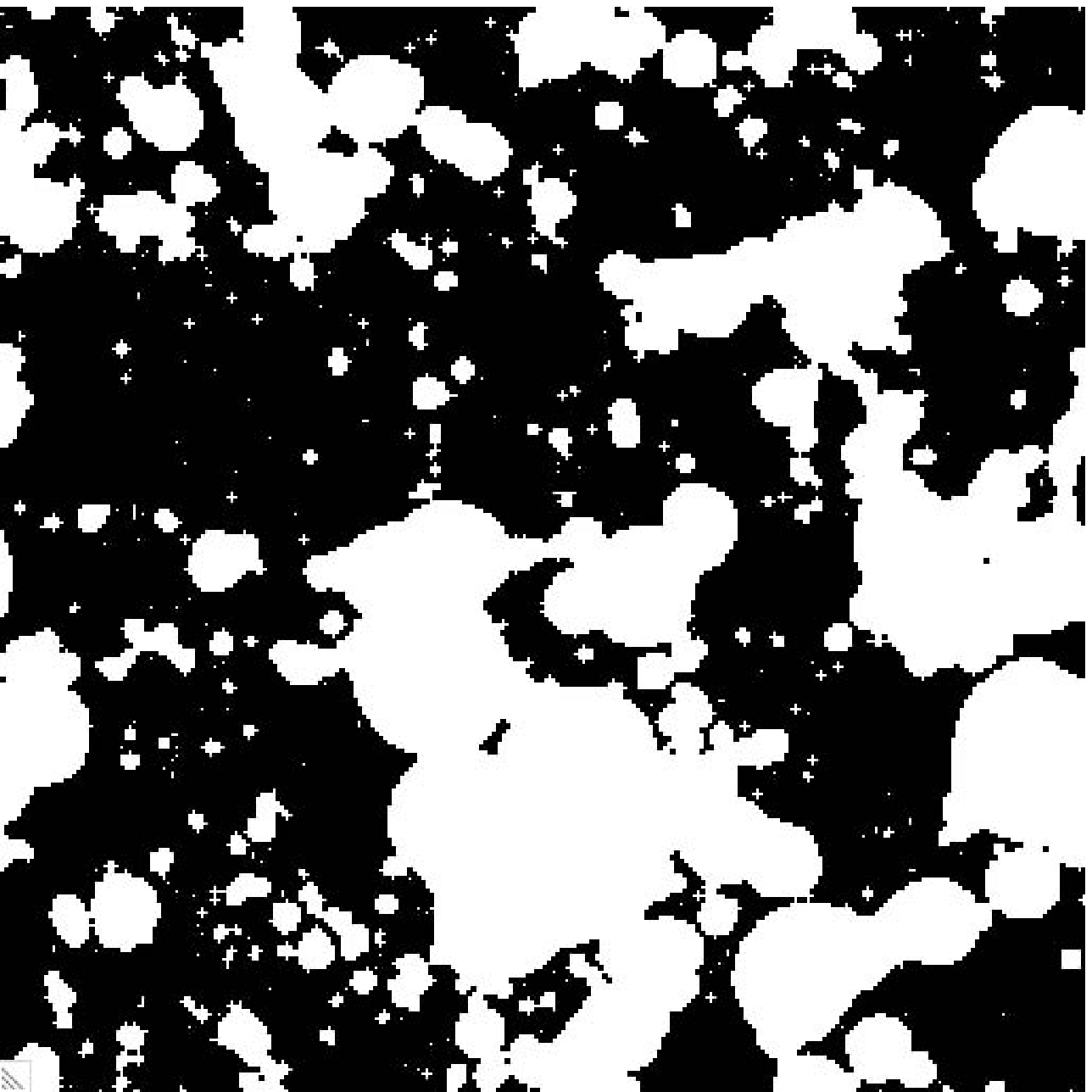}
\includegraphics[width=0.245\textwidth]{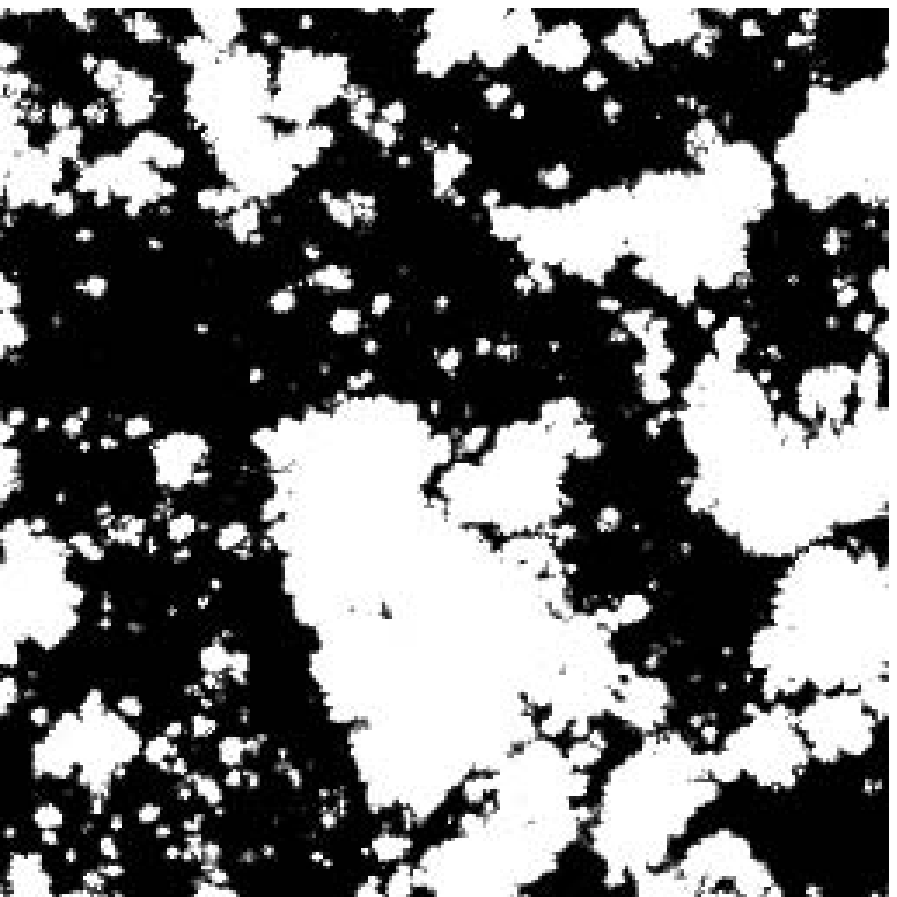}
}
\vspace{-1\baselineskip} \figcaption{
Slices from the ionization field at $z=6.89$ created using different algorithms.  All slices are 93.7 Mpc on a side and 0.37 Mpc deep, with the mean neutral fraction in the box being $\avenf=0.49$.  Ionized regions are shown as white.  The left-most panel was created by performing the bubble filtering procedure of \citet{Zahn07} directly on the linear density field.  The second panel was created by performing their bubble filtering procedure on their N-body halo field, but with the slightly different barrier definition in eq. (\ref{eq:HII_barrier}).  The third panel was created by performing our bubble filtering procedure described in \S~\ref{sec:bubble_filtering} on the same N-body halo field.  The right-most panel (from \citealt{Zahn07}) was created using an approximate RT algorithm on the same halo field.
\label{fig:oliver_cmp}
}
\vspace{-1\baselineskip}
\end{figure*}

In order to test our bubble filtering algorithm, we execute it on the same N-body halo field at $z=6.89$ as was used to generate the bottom panels of Fig. 3 in \citet{Zahn07}.  We compare analogous ionization maps created using various algorithms in Figure \ref{fig:oliver_cmp}.  All slices are 93.7 Mpc on a side and 0.37 Mpc deep, with $\zeta$ adjusted so that the mean neutral fraction in the box is $\avenf=0.49$.  Ionized regions are shown as white.  The left-most and right-most panels are taken from \citet{Zahn07}.  The left-most panel was created by performing their bubble filtering procedure directly on the linear density field (without explicitly identifying halos).  The second panel was created by performing their bubble filtering procedure on their N-body halo field, but with the slightly different barrier definition in eq. (\ref{eq:HII_barrier}).  
The third panel was created by performing our bubble filtering procedure on the same N-body halo field, but ignoring density fluctuations outside of halos (i.e. setting $\langle \deltanlxvec \rangle_R = 0$), which we have verified give nearly identical bubble maps as our full procedure (so long as $\avenf$ is fixed).  The right-most panel was created using an approximate RT algorithm \citep{AW02, SAH01, Sokasian03} on the same halo field.

It is immediately obvious from Fig. \ref{fig:oliver_cmp} that all of the approximate maps ({\it first three panels}) reproduce the RT map ({\it right-most panel}) fairly well.  Even the HII bubble filtering performed directly on the linear density field ({\it left-most panel}) performs well, which is encouraging, as that is the starting point for our semi-numerical procedure and we only improve on this scheme.

Figure \ref{fig:oliver_cmp} shows that our HII bubble filtering algorithm is an excellent approximation to RT.  The similar algorithm proposed by \citet{Zahn07} also performs well.  In comparison, our algorithm produces somewhat more ``bubbly'' maps but appears to better capture the connectivity of HII regions.  Both are an obvious improvement on directly filtering the linear density field.

\begin{figure*}
\vspace{+0\baselineskip}
{
\includegraphics[width=0.5\textwidth]{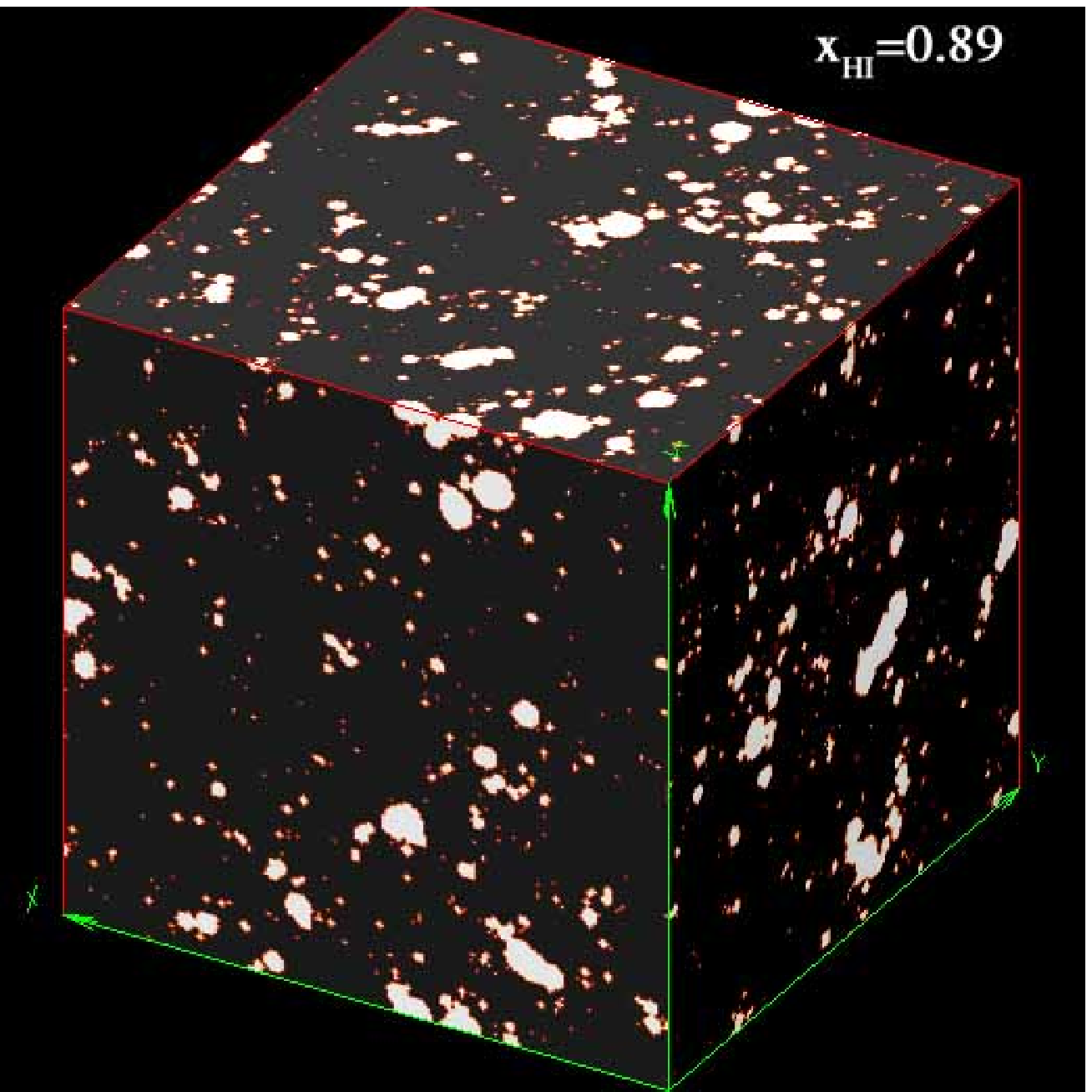}
\includegraphics[width=0.5\textwidth]{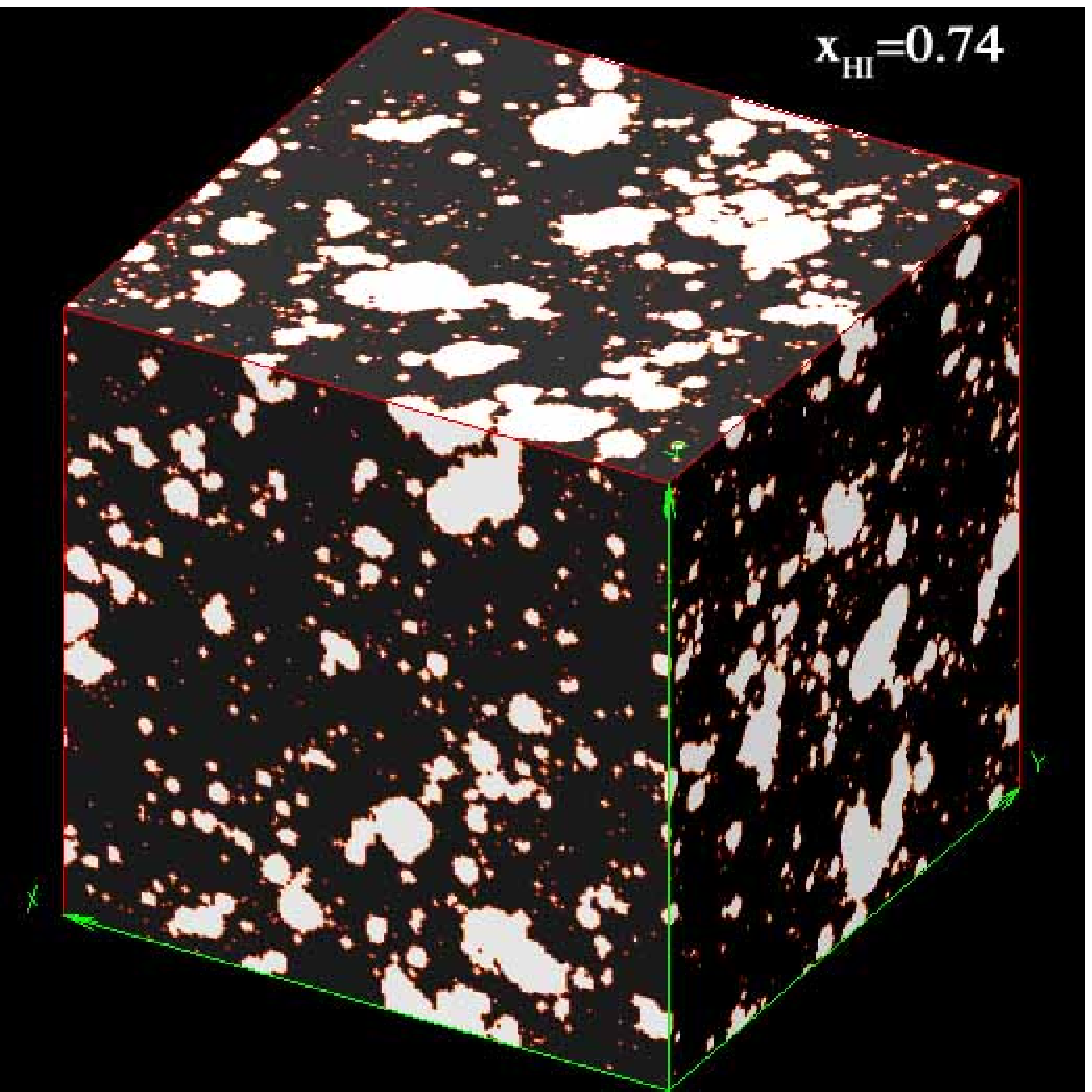}
}
{
\includegraphics[width=0.5\textwidth]{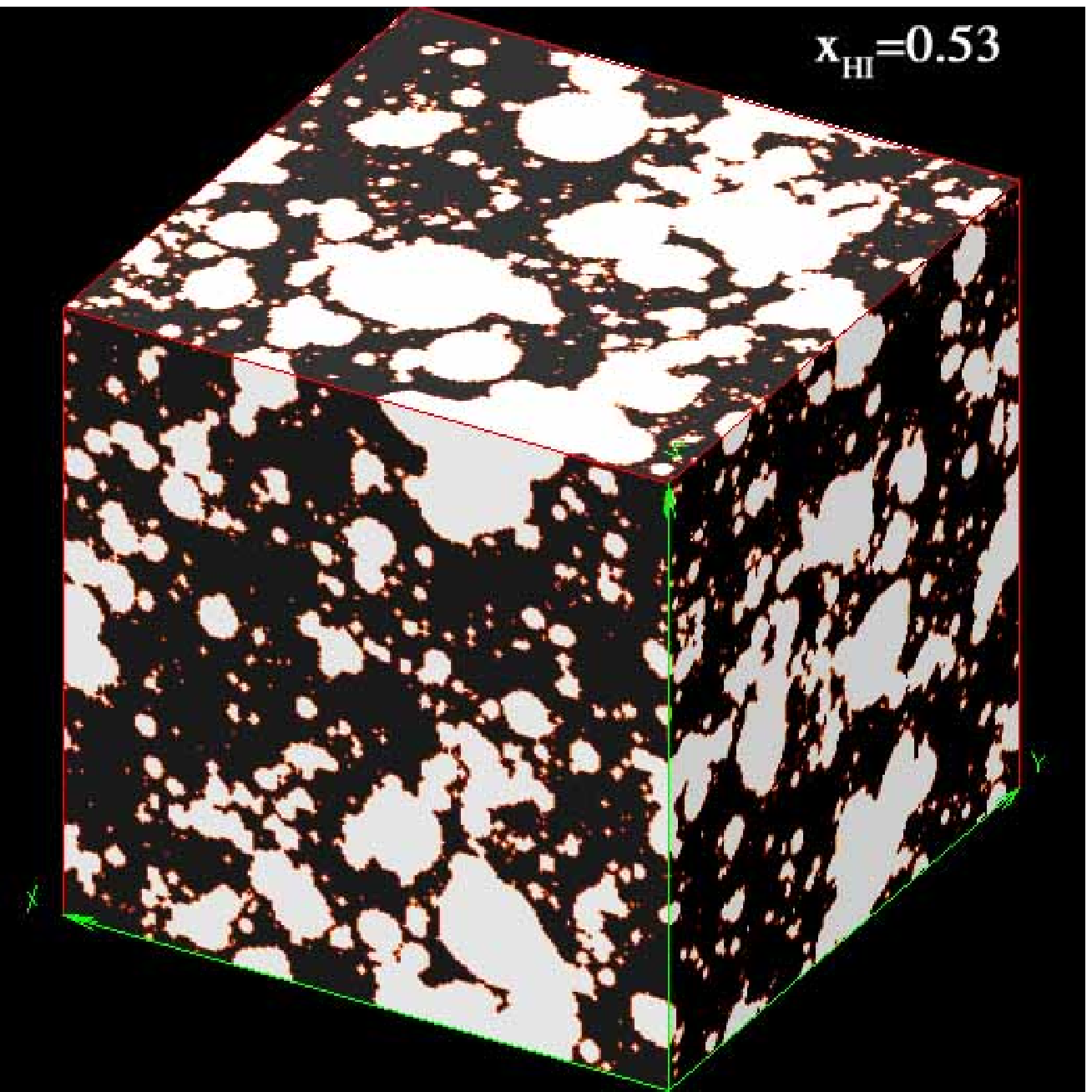}
\includegraphics[width=0.5\textwidth]{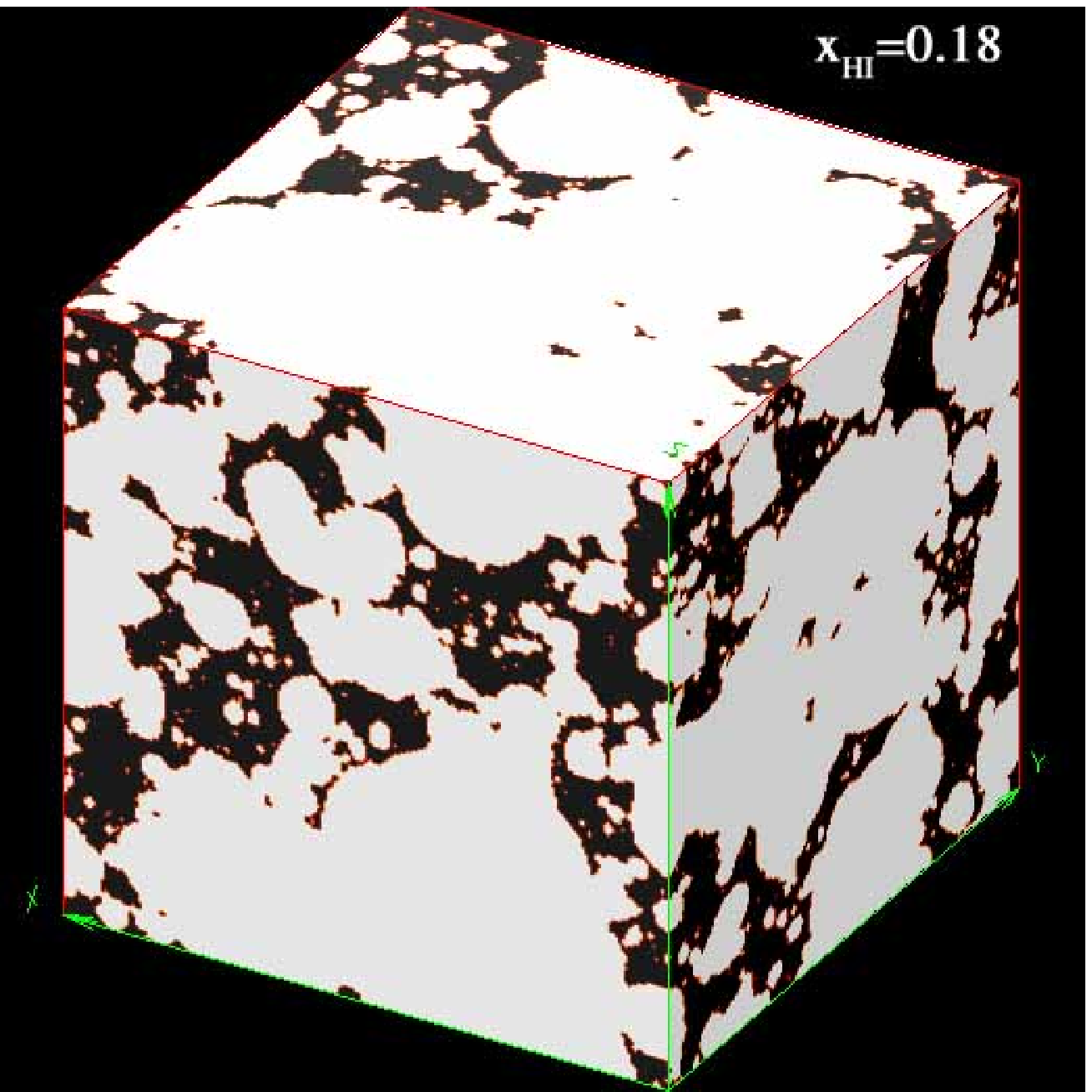}
}
\vspace{-1\baselineskip} \figcaption{
Slices through the $200^3$ ionization field at $z=$ 10, 9, 8.25, 7.25 ({\it left to right across rows}).  With the assumption of $\zeta=15.1$, these redshifts correspond to $\avenf=$ 0.89, 0.74, 0.53, 0.18, respectively.  All slices are 100 Mpc on a side and 0.5 Mpc deep.  The bottom-left panel corresponds to the halo field in the top-right panel of Fig. \ref{fig:halo_slice}, generated on a high-resolution 1200$^3$ grid.
\label{fig:bubble_maps}
}
\vspace{-1\baselineskip}
\end{figure*}

Of course, in our full algorithm we identify halos from the linear density field (rather than from simulations), so our method consumes comparable processing time to the one used to generate the leftmost panel in Figure~\ref{fig:oliver_cmp}, once the halos have been identified.  Moreover, we are able to capture the ``stochastic" component of the halo bias that causes the relatively large differences between the leftmost panel and the full RT simulation.  That is, the algorithm used to generate the leftmost panel uses the large-scale linear density field to \emph{predict} the distribution of halos \citep{Zahn05, Zahn07}.  In reality, the relation is not deterministic because of random fluctuations in the small-scale modes comprising each region.  This leads to nearly Poisson scatter in the halo number densities \citep{SL99, Casas-Miranda02} that can substantially modify the bubble size distribution whenever sources are rare, particularly early in reionization \citep{FMH05}.  By directly sampling the small-scale modes to build the halo distribution, we better recover this scatter (at least statistically, as illustrated by Fig.~\ref{fig:halo_ps}).  Another way to include this scatter is by directly sampling halos from an N-body simulation (as in \citealt{Zahn07}, or the second panel of Fig.~\ref{fig:oliver_cmp}), although that obviously requires much more computing power.

\subsection{Ionization Maps}
\label{sec:maps}

Now that we have demonstrated in turn the success of our halo and bubble filtering procedures, we present the resulting ionization maps when the two are combined.  In Figure \ref{fig:bubble_maps}, we show 100 Mpc $\times$ 100 Mpc $\times$ 0.5 Mpc slices through our $200^3$ ionization field at $z=$ 10, 9, 8.25, 7.25 ({\it left to right across rows}).  With the assumption of $\zeta=15.1$, these redshifts correspond to $\avenf=$ 0.89, 0.74, 0.53, 0.18, respectively.  As has been pointed out by \citet{FZH04}, the neutral fraction is the more relevant descriptor; bubble morphologies at a constant $\avenf$ vary little with redshift (see also \citealt{McQuinn06}).  The bottom-left panel corresponds to the halo field in the top-right panel of Fig. \ref{fig:halo_slice}, generated on a high-resolution 1200$^3$ grid.

\begin{figure}
\vspace{+0\baselineskip}
\myputfigure{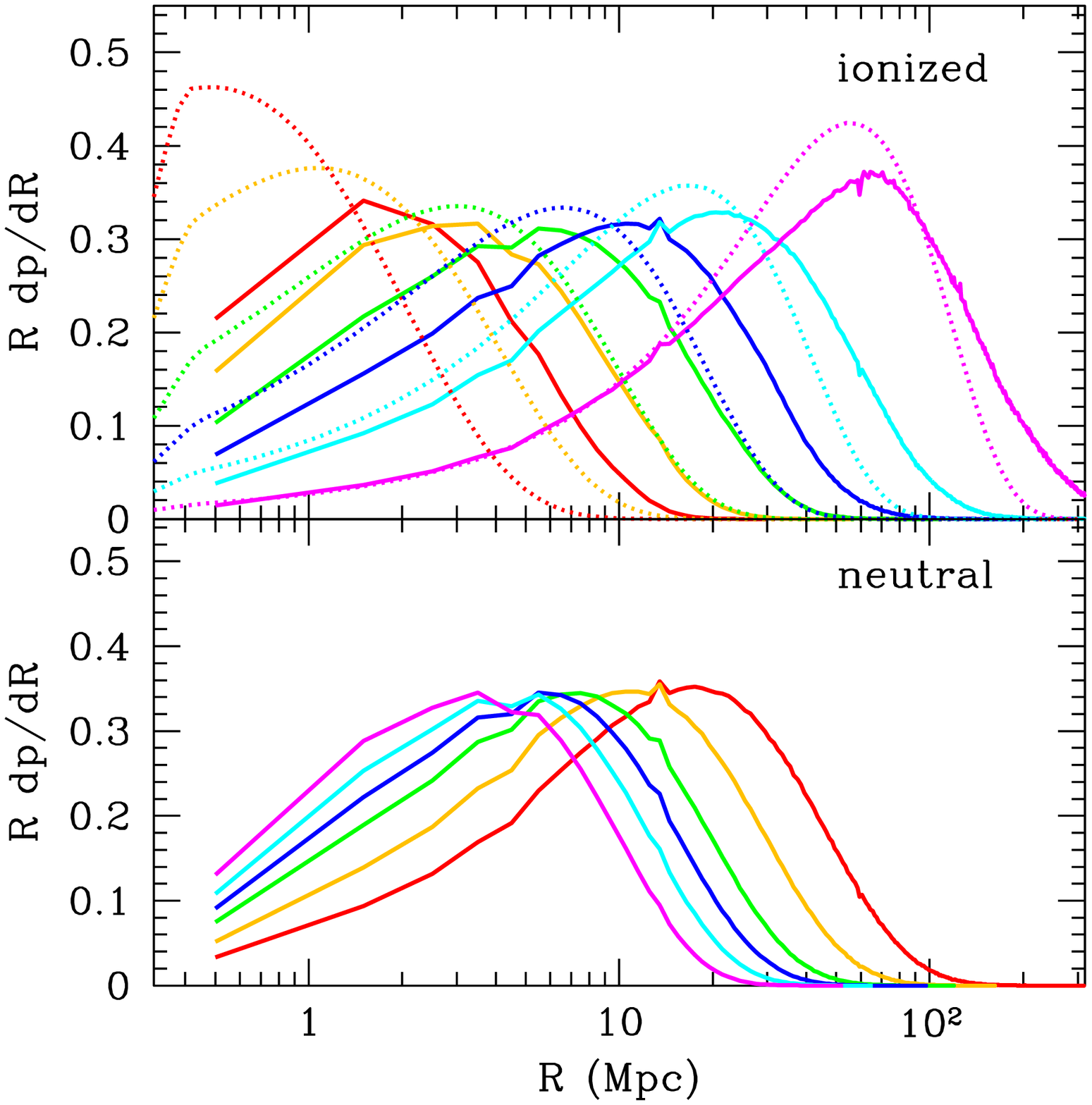}{3.3}{0.5}{.}{0.}  \figcaption{ 
Size distributions (see definition in text) of ionized ({\it top panel}) and neutral ({\it bottom panel}) regions.  Curves correspond to ($z$, $\avenf$) = (10, 0.89), (9.25, 0.79), (8.50, 0.61), (8.00, 0.45), (7.50, 0.27), (7.00, 0.10), from left to right in the top panel, respectively (or from right to left in the bottom panel).  Solid curves are produced from our simulation while dotted curves correspond to the analytic mass function.  All curves are normalized so that the probability distribution integrates to unity.
\label{fig:bubble_pdf}}
\vspace{-1\baselineskip}
\end{figure}

To quantify the ionization topology resulting from our method, we calculate the size distributions of both the ionized and neutral regions.  We randomly choose a pixel of the desired phase (neutral or ionized), and record the distance from that pixel to a phase transition along a randomly chosen direction.  We repeat this Monte Carlo procedure 10$^7$ times.  Volume-weighted probability distribution functions (PDFs) produced thusly are shown by the solid curves in Figure \ref{fig:bubble_pdf} for ionized regions ({\it top panel}) and neutral regions ({\it bottom panel}). Curves correspond to ($z$, $\avenf$) = (10, 0.89), (9.25, 0.79), (8.50, 0.61), (8.00, 0.45), (7.50, 0.27), (7.00, 0.10), from left to right in the top panel, respectively (or from right to left in the bottom panel).  All curves are normalized so that the probability density integrates to unity.  

It is useful to compare these distributions to the analytic bubble mass function of \citet{FZH04}; although this analytic approach is motivated by the same excursion set barriers as our semi-numerical approach, it does not account for the full geometry of sources.  We compute the probability distribution from the analytic model by assuming purely spherical bubbles and convolving with the volume-weighted distance to the sphere's edge:
\begin{equation}
p(r) \, dr= {2 \pi r^2 \, dr \over (1-\avenf)} \int dR \, n_b(R) \left( 1 - {r \over 2 R} \right),
\end{equation}
where $n_b(R)$ is the comoving number density of bubbles with radii between $R$ and $R+dR$ (taken from \citealt{FZH04}).

  Several points are evident from Figures \ref{fig:bubble_maps} and \ref{fig:bubble_pdf}.  As expected (e.g., \citealt{FZH04, FMH05, McQuinn06}), there is a well-defined bubble scale at each neutral fraction, despite some scatter in the sizes.  This scale also gets more pronounced (i.e. the PDF peaks 
more) as reionization progresses; this is a result of the changing shape of the underlying matter power spectrum \citep{FMH05}.

  Also, the purely analytic estimates underpredict the size distributions at all values of the neutral fraction, though they do become increasingly accurate as the neutral fraction decreases.  This trend is perhaps counterintuitive, as the analytic model, which rests on the assumption of spherical bubbles, should perform best when the bubbles are isolated, as one would expect at earlier times, i.e. high neutral fractions.  However, looking at the top-left panel of Fig. \ref{fig:bubble_maps}, the typical bubbles filling most of the ionized volume overlap due to the strong clustering of early sources and bubbles.  This results in many ``overlapping pairs of spheres'' at early times, resulting from merging HII bubbles sourced by clustered sources.  Thus the {\it spherical} bubble-based analytic model underpredicts the true size distribution, using our ``mean free path'' definition of bubble sizes above.  
This effect was not noted by previous studies \citep{Zahn07}, because they used a different definition of bubble sizes, based on spherical filters used to flag regions in which $\avenf < 0.1$.  As time progresses and the universe becomes more ionized, this ``overlapping pair of spheres'' effect becomes less and less dominant (see Fig. \ref{fig:bubble_maps}), and the analytic model becomes increasingly more accurate.

Finally, the size distributions of neutral regions presented in the bottom panel of Fig. \ref{fig:bubble_pdf} are a new result and potentially important for the 21-cm signal (which originates in neutral hydrogen, of course).  In the later stages of reionization, when the topology has transformed to isolated neutral islands in a sea of ionized gas, this figure pinpoints the typical sizes of ``mostly neutral" pixels that continue to emit strongly.  In contrast to the ionized regions, the neutral regions (defined in this way) do not grow substantially during reionization.  From $\avenf=0.89$ to $\avenf=0.1$, the peak of the distribution shifts only by a factor of $\sim$ 6, whereas the peak of the ionized region distribution shifts by a factor of $\sim$40 over the same range.  The reason for this is also evident in Figure \ref{fig:bubble_maps}: even when the universe is mostly neutral, space is dotted with islands of ionized gas, such that our ``mean free path''--type size distributions never become too large.  The converse does not hold true for ionized regions.  However, a slight parallel for ionized islands in a mostly neutral IGM, could be found in Lyman limit systems (LLS) inside larger HII regions (e.g. \citealt{BL02_screening, SIR04, MeHR00}), though it is not clear how prevalent such neutral clumps are at high redshifts.

\begin{figure}
\vspace{+0\baselineskip}
\myputfigure{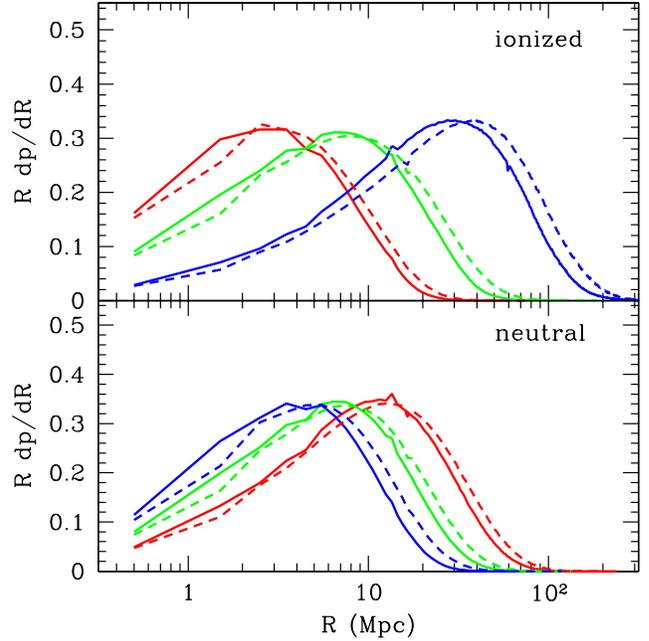}{3.3}{0.5}{.}{0.}  \figcaption{ 
Size distributions of ionized ({\it top panel}) and neutral ({\it bottom panel}) regions from different simulation boxes.  Curves correspond to ($z$, $\avenf$) = (9.00, 0.80), (8.00, 0.56), (7.00, 0.21), from left to right in the top panel, respectively (or from right to left in the bottom panel).  Solid curves are generated from our fiducial, $N=1200^3$, $L=$ 100 Mpc, simulation while dashed curves are generated from a larger simulation with $N=1500^3$, $L=$ 250 Mpc.  The cell size in all ionization maps is 0.5 Mpc on a side, with the efficiency parameter, $\zeta$, adjusted to get matching values of $\avenf$ and the minimum halo mass set to $M_{\rm min} = 2.2 \times 10^8 \Msun$ for comparison purposes.
\label{fig:bubble_pdf_difflen}}
\vspace{-1\baselineskip}
\end{figure}

Throughout this paper, we have used a $L=100$ Mpc ``simulation'' box.  This size facilitates comparison of our results with those from recent hybrid N-body works \citep{Zahn07, McQuinn06, Iliev06, TC06}.  However, the speed of our semi-numerical approach allows us to explore larger cosmological scales while still {\it consistently} resolving the small halos that could dominate the photon budget during reionization.  As mentioned previously, existing N-body codes must resort to merger-tree methods to populate their distribution of small-mass halos, even for box sizes $\lsim$ 100 Mpc \citep{McQuinn06}.  In this spirit, we present some preliminary results from a $N=1500^3$, $L=$ 250 Mpc simulation, capable of directly resolving halos with masses $M \gsim 2.2 \times 10^8 \Msun$, with resulting mass functions accurate to better than a factor of two even at the smallest scale.  This resolution pushes the RAM limit of our machine and so each redshift can take several hours to complete.\footnote{We note here that our halo-finding algorithm requires significantly higher resolution than does predicting the ionization field directly from the linear density field smoothed on larger scales \citep{Zahn05,Zahn07}.  The latter method can be extended to even larger boxes, though at the price of a  somewhat less accurate ionization map (compare the left and right panels in Fig.~\ref{fig:oliver_cmp}).}  

In Figure \ref{fig:bubble_pdf_difflen}, we compare size distributions of ionized ({\it top panel}) and neutral ({\it bottom panel}) regions from our two different simulation boxes.  Curves correspond to ($z$, $\avenf$) = (9.00, 0.80), (8.00, 0.56), (7.00, 0.21), from left to right in the top panel, respectively (or from right to left in the bottom panel, respectively).  Solid curves are generated from our fiducial, $N=1200^3$, $L=$ 100 Mpc, simulation while dashed curves are generated from our larger simulation with $N=1500^3$, $L=$ 250 Mpc.  The cell size in all ionization maps is 0.5 Mpc on a side, with the efficiency parameter, $\zeta$, adjusted to obtain matching values of $\avenf$, and we set the minimum halo mass to $M_{\rm min} = 2.2 \times 10^8 \Msun$ even in the higher resolution runs for easier comparison.

As reionization progresses, an increasing number of large HII regions are ``missed'' by the $L=$ 100 Mpc simulation.  Interestingly, the analogous trend in the neutral region size distributions ({\it bottom panel}) is weaker.  This is most likely because the ``ionized island'' effect limits the size distributions of neutral regions as described above.

\section{21-cm Temperature Fluctuations}
\label{sec:21cm}

A natural application of our ``simulation'' technique is to predict 21-cm brightness temperatures during reionization.  The offset of the 21-cm brightness temperature from the CMB temperature, $\Tcmb$, along a line of sight (LOS) at observed frequency $\nu$, can be written as (e.g. \citealt{FOB06}):

\begin{eqnarray}
\label{eq:delT}
\delT(\nu) &=& \frac{T_S - \Tcmb}{1+z} (1 - e^{-\tau_{\nu_0}}) \\
\nonumber &\approx& 9 (1+z)^{1/2} x_{\rm HI} (1+\delNL) \frac{H}{dv_r/dr + H} ~~ {\rm mK},
\end{eqnarray}

\noindent where $T_S$ is the gas spin temperature, $\tau_{\nu_0}$ is the optical depth at the 21-cm frequency $\nu_0$, $\delNL$ is the physical overdensity (see discussion under eq. \ref{eq:HII_barrier}), $H$ is the Hubble parameter, $dv_r/dr$ is the comoving gradient of the line of sight component of the comoving velocity, and all quantities are evaluated at redshift $z=\nu_0/\nu - 1$.  The final approximation makes the standard assumption that $T_S \gg \Tcmb$ for all redshifts of interest during reionization (e.g. \citealt{Furlanetto06}) and also that $dv_r/dr \ll H$.  We verify in our simulation that $dv_r/dr < H$ for all neutral pixels.


\begin{figure*}
\vspace{+0\baselineskip}
{
\includegraphics[width=0.33\textwidth]{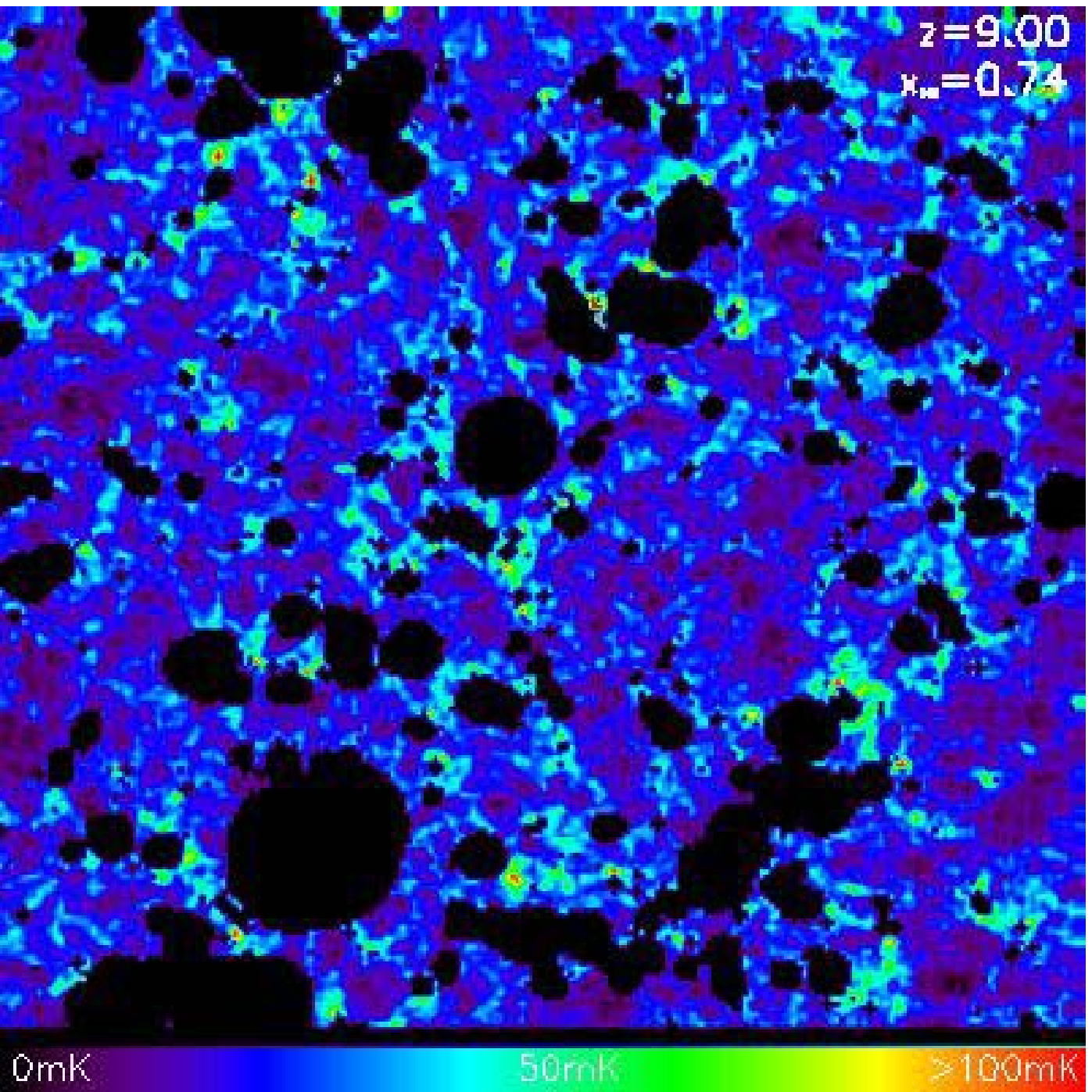}
\includegraphics[width=0.33\textwidth]{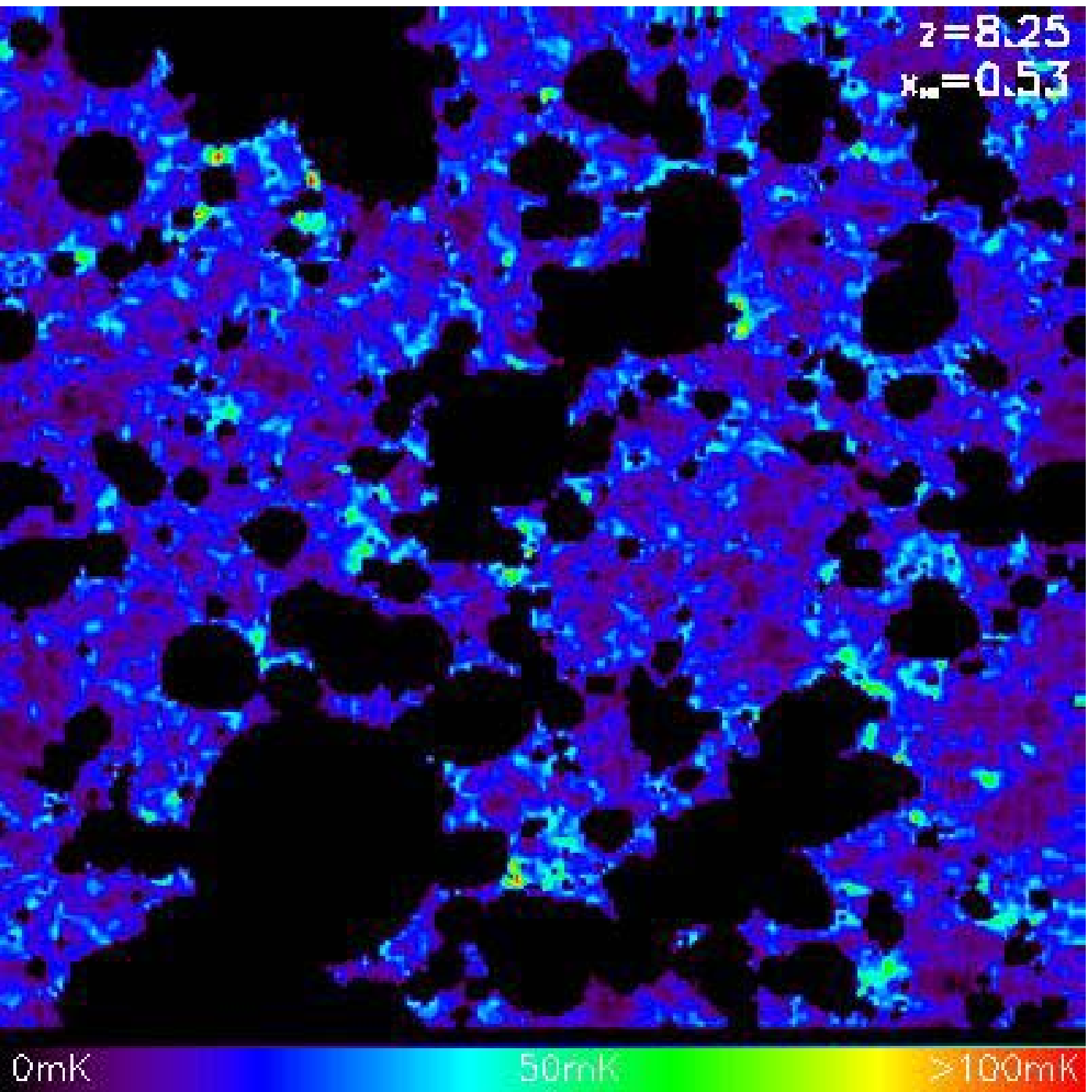}
\includegraphics[width=0.33\textwidth]{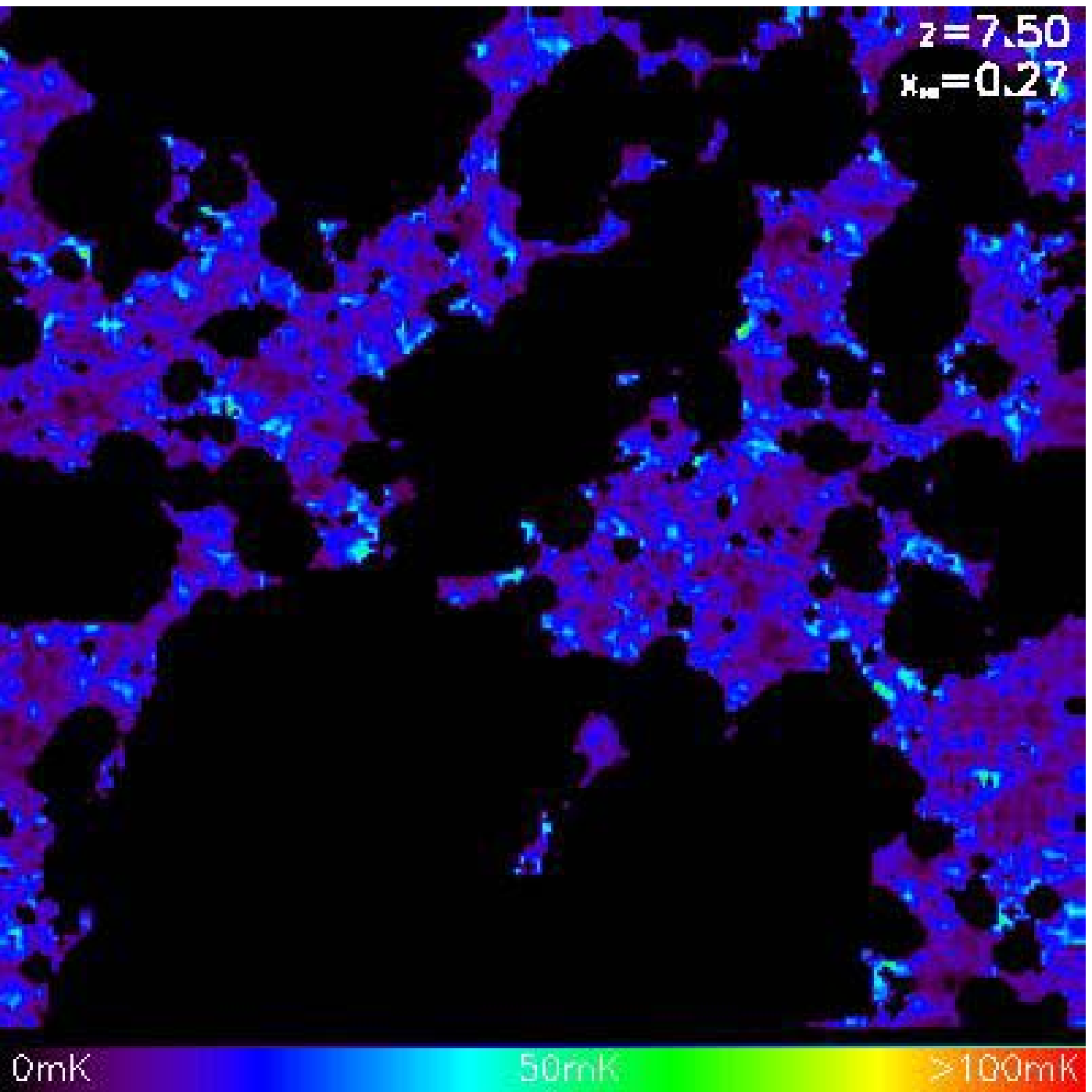}
}
{
\includegraphics[width=0.33\textwidth]{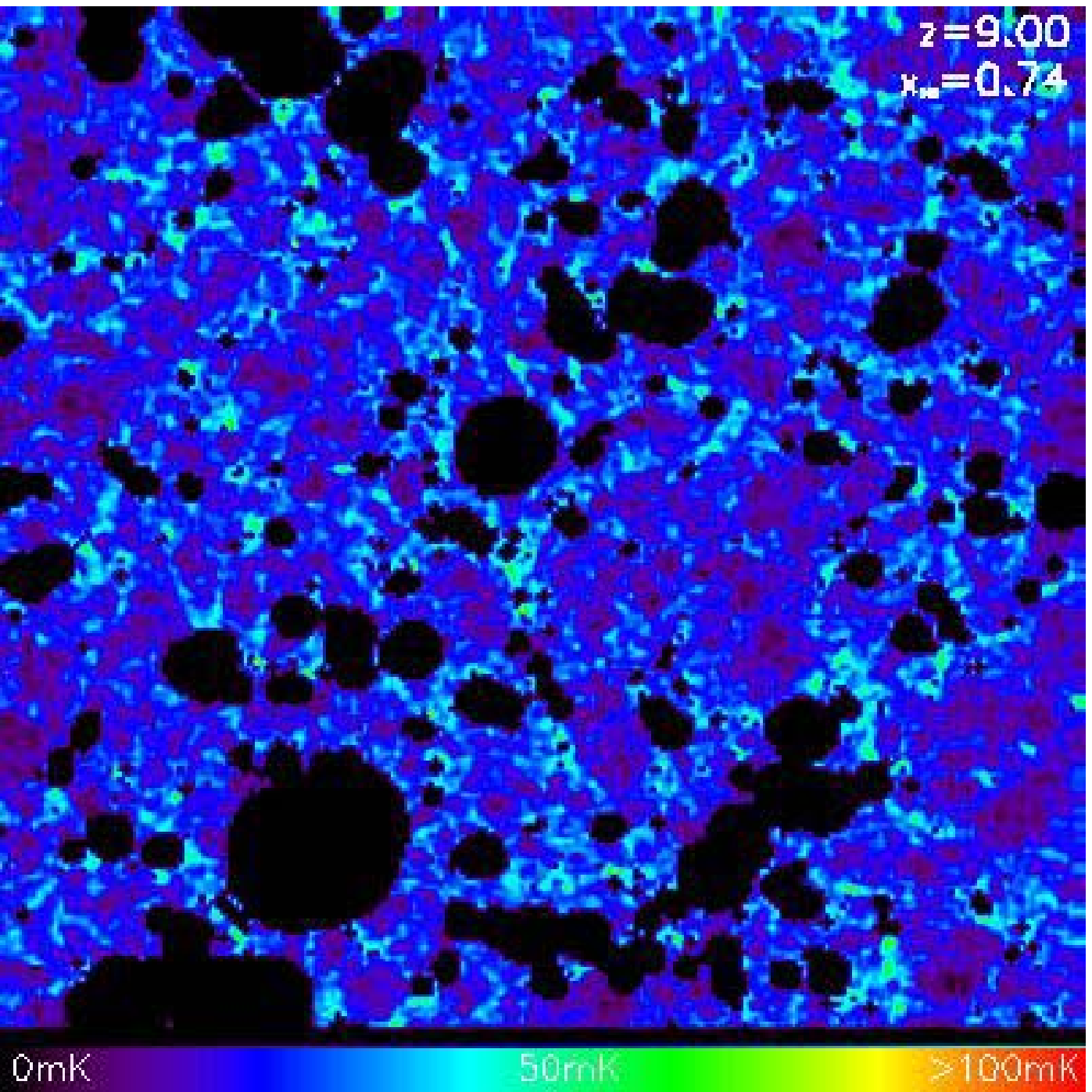}
\includegraphics[width=0.33\textwidth]{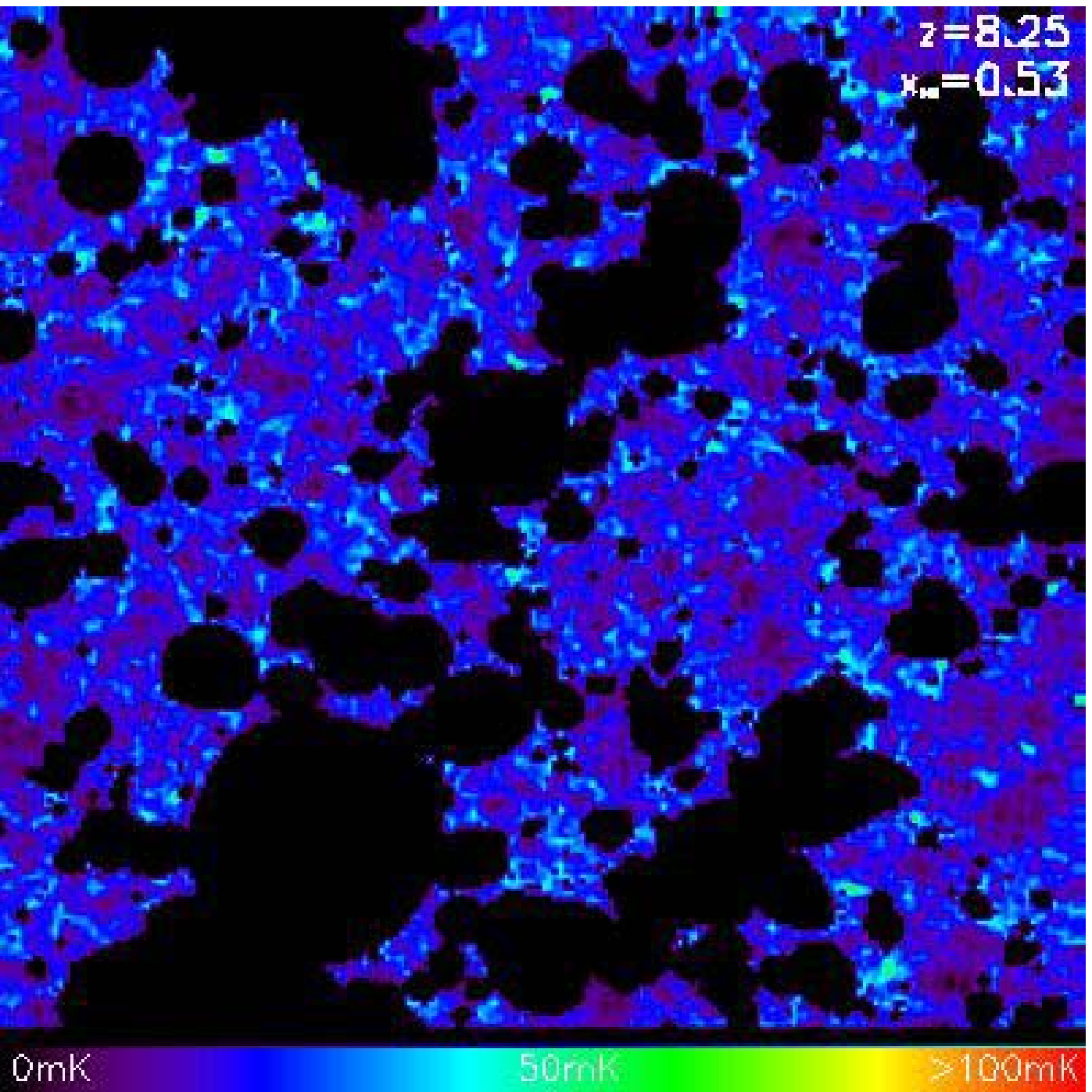}
\includegraphics[width=0.33\textwidth]{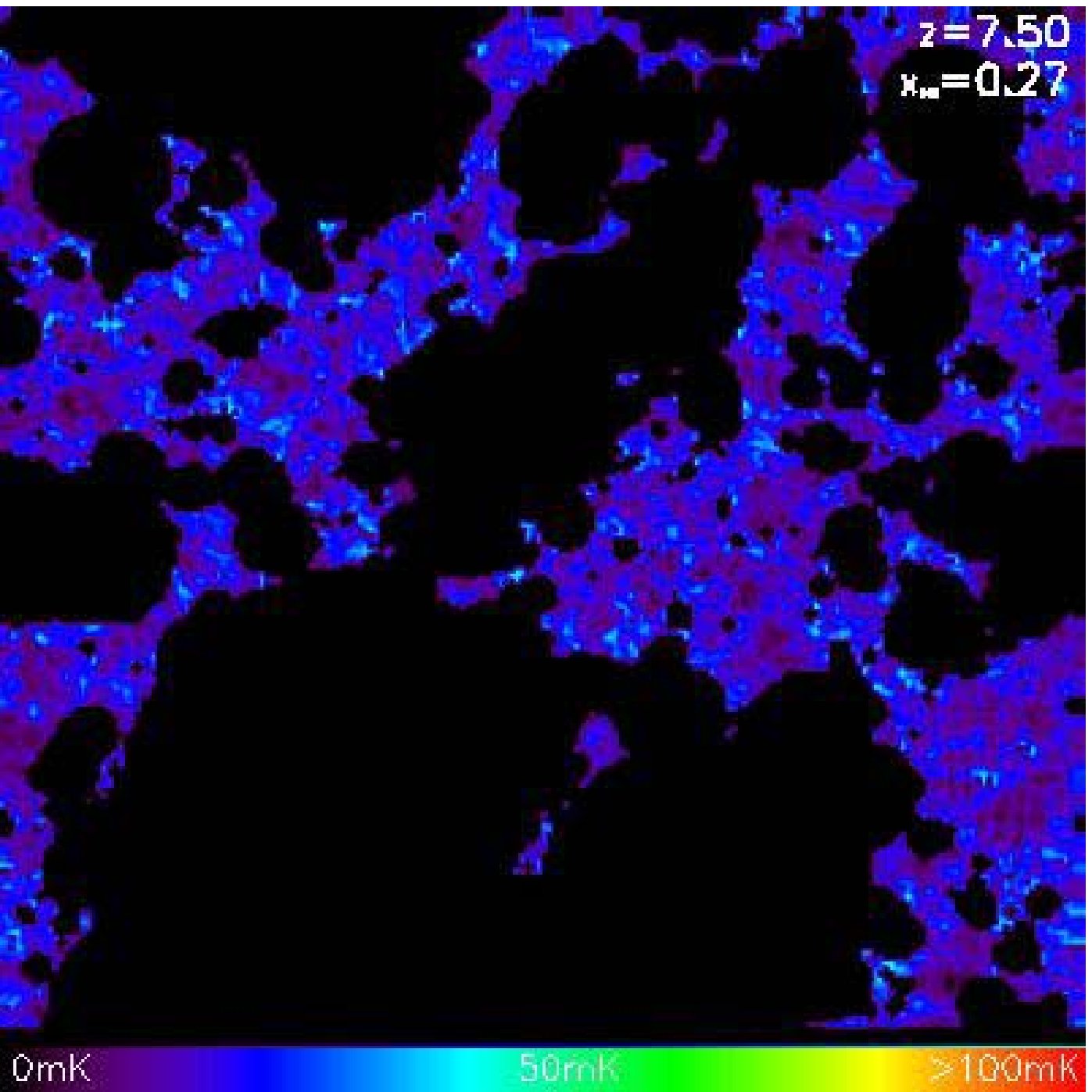}
}
\vspace{-1\baselineskip} \figcaption{
Brightness temperature of 21-cm radiation relative to the CMB temperature. All slices are 100 Mpc on a side, 0.5 Mpc deep, and correspond to ($z$, $\avenf$) = (9.00, 0.74), (8.25, 0.53), (7.50, 0.27), left to right.  Top panels include the velocity correction term in eq. (\ref{eq:delT}), while the bottom panels do not.  For animated versions of these pictures, see http://pantheon.yale.edu/$\sim$am834/Sim.
\label{fig:delta_T_maps}
}
\vspace{-1\baselineskip}
\end{figure*}

Maps of $\delT({\bf x}, \nu)$ generated in this manner are shown in Figure \ref{fig:delta_T_maps}.  All slices are 100 Mpc on a side, 0.5 Mpc deep, and correspond to ($z$, $\avenf$) = (9.00, 0.74), (8.25, 0.53), (7.50, 0.27), from left to right.  The top panels take into account the velocity correction term in eq. (\ref{eq:delT}), while the bottom panels ignore it.

As seen in Fig. \ref{fig:delta_T_maps}, velocities typically increase the contrast in temperature maps, making hot spots hotter and cool spots cooler.  We also see that temperature hot spots, which correspond to dense pixels, tend to cluster around the edges of HII bubbles, especially smaller bubbles.  This occurs because HII bubbles correlate with peaks of the density field and long-wavelength biases in the density field can extend beyond the edge of the ionized region.  This enhanced contrast might be useful in the detection of the boundaries of ionized regions with future 21-cm experiments.  As reionization progresses most hot spots become swallowed up by HII bubbles, and the effects of velocities diminish.

\begin{figure}
\vspace{+0\baselineskip}
\myputfigure{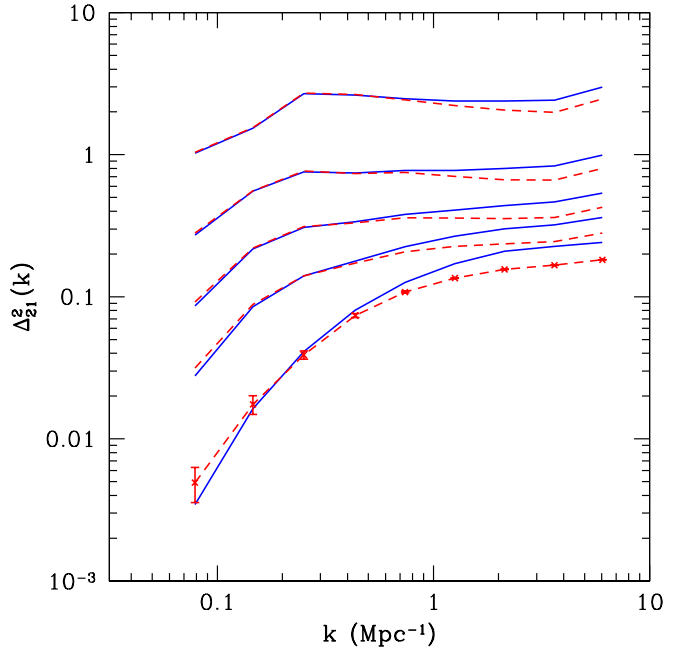}{3.3}{0.5}{.}{0.}  \figcaption{
Dimensionless 21-cm power spectra for ($\avenf$, z) = (0.79, 9.25), (0.61, 8.50), (0.45, 8.00), (0.27, 7.50), (0.10, 7.00), {\it bottom to top}.  Solid blue curves take into account gas velocities, while dashed red curves do not.
\label{fig:deltaT_ps}}
\vspace{-1\baselineskip}
\end{figure}

In Figure \ref{fig:deltaT_ps} we plot the {\it dimensionless} 21-cm power spectrum, defined as $\Delta^2_{21}(k, z) = k^3/(2\pi^2 V) ~ \langle|\delta_{\rm 21}({\bf k}, z)|^2\rangle_k$, where $\delta_{21}({\bf x}, z) \equiv \delT({\bf x}, z)/ \bar{\delT}(z) - 1$.  Solid blue curves take into account gas velocities, while dashed red curves do not.  Curves correspond to ($\avenf$, $z$) = (0.79, 9.25), (0.61, 8.50), (0.45, 8.00), (0.27, 7.50), (0.10, 7.00), {\it bottom to top}.
Error bars on the bottom dashed curve denote 1-$\sigma$ Poisson uncertainties; fractional errors in a given bin are the same for all curves.
  As reionization progresses, small-scale power is traded for large-scale power, and the curves become flatter.  
Note that, with our dimensionless definition of the power spectrum, curves with smaller $\avenf$ have \emph{larger} values of $\Delta^2_{21}(k, z)$.  This is because the mean brightness temperature offset drops quite rapidly as reionization progresses, since $\bar{\delT}(z) \propto \avenf$, but the scatter remains significant (see Fig.~\ref{fig:bubble_pdf}) and thus the {\it fractional} perturbation, $\delta_{21}({\bf x}, z)$, increases throughout reionization.

\begin{figure}
\vspace{+0\baselineskip}
\myputfigure{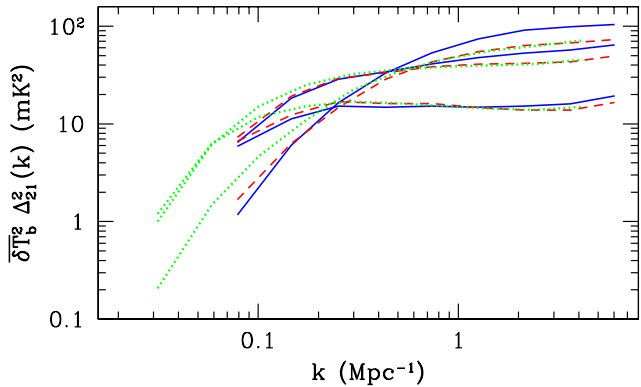}{3.3}{0.5}{.}{0.}  \figcaption{ 
Dimensional 21-cm power spectra.  The curves correspond to ($\avenf$, z) = (0.80, 9.00), (0.56, 8.00), (0.21, 7.00), top to bottom at large $k$, respectively.
  The dotted green curves are generated from a large, high-resolution ``simulation'', with $N=1500^3$ and $L = 250$ Mpc, with no velocity contribution to the power spectra.  Solid blue curves and dashed red curves are generated with our fiducial $N=1200^3$ and $L = 100$ Mpc simulation, with and without the velocity contribution, respectively. 
\label{fig:ps_difflen}}
\vspace{-1\baselineskip}
\end{figure}

Finally, in Figure \ref{fig:ps_difflen} we plot {\it dimensional} power spectra, $\bar{\delT}(z)^2 \Delta^2_{21}(k, z)$.  The curves correspond to ($\avenf$, $z$) = (0.80, 9.00), (0.56, 8.00), (0.21, 7.00), top to bottom at large $k$, respectively.  The dotted green curves are generated from a large, high-resolution ``simulation'', with $N=1500^3$ and $L = 250$ Mpc, with no velocity contribution to the power spectra.  Solid blue curves and dashed red curves are generated with our fiducial $N=1200^3$ and $L = 100$ Mpc simulation, with and without the velocity contribution, respectively.  The cell size in all $\delT$ maps is 0.5 Mpc on a side, with the efficiency parameter, $\zeta$, adjusted to achieve matching values of $\avenf$ and the minimum halo mass set to $M_{\rm min} = 2.2 \times 10^8 \Msun$ for comparison purposes.

As seen in Figures \ref{fig:deltaT_ps} and \ref{fig:ps_difflen}, velocities make a modest contribution to the 21 cm power spectrum, boosting power on small scales early in reionization.  Note that the apparent slight decrease in power at small $k$ when velocities are included is well within the errors from averaging over the few modes available to us on the largest scales (e.g., see Poisson error bars on the bottom dashed curve in Fig. \ref{fig:deltaT_ps}).
 While the maximum $\delT$ value in our simulation box increases by a factor of a few when velocities are included, most of the pixels are only slightly affected.  When the power spectrum is plotted in a dimensional version, $\bar{\delT}(z)^2 \Delta^2_{21}(k, z)$, small-scale power is boosted by $\sim40\%$
 at ($\avenf$, $z$) = (0.80, 9.00), with this enhancement monotonically decreasing as reionization progresses.   Linear theory predicts that velocities enhance the \emph{density} power spectrum by a factor of 1.87 when $\avenf=1$ \citep{Kaiser87}.  In fact we do recover this enhancement for a fully neutral IGM; however, as predicted by analytic models \citep{McQuinn06_21cm}, the ionized bubbles rapidly remove most of this amplification.

Figure \ref{fig:ps_difflen} also confirms the inferences drawn from Fig. \ref{fig:bubble_pdf_difflen}, primarily that larger box sizes are needed to capture the ionization topology at the end stages of reionization.  Comparing the dashed red to the dotted green curves in Fig. \ref{fig:ps_difflen}, we note that our fiducial $L=$ 100 Mpc simulations are accurate for scales smaller than $k\gsim0.2$ Mpc$^{-1}$ (or $\lambda \la 30$ Mpc).  As reionization progresses, larger scales lose power more rapidly than in the $L=$ 250 Mpc simulation.  This is again evidence that very large scale simulations are needed to model the middle and late stages of reionization.  Thus the speed and high resolution of our semi-numeric approach will be extremely useful in future modeling of reionization.

\section{Conclusions}
\label{sec:conc}

We introduce a method to construct semi-numeric simulations that can efficiently generate realizations of halo distributions and ionization maps at high redshifts.  Our procedure combines an excursion-set approach with first-order Lagrangian perturbation theory and operates directly on the linear density and velocity fields.  As such, our algorithm can exceed the dynamic range of existing N-body codes by orders of magnitude.  As this is the main limiting factor in simulating the ionized bubble topology throughout reionization, when ionized regions reach scales of tens of comoving Mpc, this will be particularly useful in such studies.  Moreover, the efficiency of the algorithm will allow us to explore the large parameter space required by the many uncertainties associated with high-redshift galaxy formation.

We find that our halo finding algorithm compares well with N-body simulations on the statistical level, yielding both accurate mass functions and power spectra.  We have not yet compared our halo distribution with simulations on a point-by-point basis, but we do not expect perfect agreement because of the vagaries of the excursion set approach.  However, it is encouraging that a very similar algorithm independently developed by \citet{BM96_algo} fares quite well in a comparison of high-mass halos.

Our HII bubble finding algorithm captures the bubble topology quite well, as compared to ionization maps from ray-tracing RT algorithms at an identical $\avenf$.  Our algorithm is similar to other codes, although we build the ionization map from our excursion set halo field rather than directly from the linear density field or from halos found in an N-body simulation \citep{Zahn05, Zahn07}.  
Compared to codes built only from the linear density field, we can better track the ``stochastic" component of the bias, though at the cost of somewhat more computation and a harder limit on resolution.  On the other hand, our scheme is much faster than using an N-body code and offers superior dynamic range.

We create ionization maps using a simple efficiency parameter and compute the size distributions of ionized and neutral regions.  Our size distributions are generally shifted to larger scales when compared with purely analytic models \citep{FZH04} at the same mean neutral fraction.  The discrepancy lies in the fact that, at their core, the purely analytic models are based on ensemble-averaged distributions of isolated spheres.  Hence they do not capture overlapping bubble shapes, which are most important at large $\avenf$ (when the bubbles are small and random fluctuations in the source densities, as well as clustering, are most important).   

In this paper, we have confined ourselves to a simple ionization criterion (essentially photon counting; \citealt{FZH04}).  However, our algorithm can easily accommodate more sophisticated prescriptions, so long as they can be expressed either with the excursion set formalism \citep{FO05, FMH05} or built from the halo field (in a similar way to semi-analytic models of galaxy formation embedded in numerical simulations).

 We also use our procedure to generate maps and power spectra of the 21-cm brightness temperature fluctuations during reionization.  We note that temperature hot spots generally cluster around HII bubbles, especially in the early phases of reionization.  Because HII bubbles correlate with peaks of the density field, long-wavelength biases in the density field can extend beyond the edge of the ionized region, with the resulting overdensities appearing as hot spots.  This effect might be useful for detecting the boundaries of ionized regions with future 21-cm experiments.
We study the imprint of gas bulk velocities on 21-cm maps and power spectra, an effect which was not included in previous studies.
   We find that velocities do not have a major impact during reionization, although they do increase the contrast in temperature maps, making some hot spots hotter and some cool spots cooler.  Velocities also increase small-scale power, though the effect decreases with decreasing $\avenf$.

We also include some preliminary results from a simulation run with the largest dynamical range to date: a 250 Mpc box which resolves halos with masses $M\gsim2.2\times10^8\Msun$.  This simulation run confirms that extremely large scales are required to model the late stages of reionization, $\avenf\lsim0.5$, when the typical scale of ionized bubbles becomes several tens of Mpc.

 The speed and dynamic range provided by our semi-numeric approach will be extremely useful in the modeling of early structure formation and reionization.  Our ionization maps can be efficiently folded into analyses of current and upcoming high-redshift observations, especially 21-cm surveys.

\acknowledgments{
We thank Greg Bryan for many helpful conversations concerning the inner workings of cosmological simulations and the generation of initial conditions.  We also thank Oliver Zahn for permitting the use of the halo field from his simulation output as well as for several interesting discussions.  We thank Mathew McQuinn for providing the halo power spectra from his simulation as well as for associated helpful comments.   We thank Zoltan Haiman, Greg Bryan, Oliver Zahn and Mathew McQuinn for insightful comments on a draft version of this paper. This research was supported by NSF-AST-0607470.
}

\bibliographystyle{apj}
\bibliography{ms}

\end{document}